\documentclass[graybox]{svmult}

\usepackage{type1cm}        
\usepackage{makeidx}         
\usepackage{graphicx}        
\usepackage{multicol}        
\usepackage[bottom]{footmisc}

\usepackage{newtxtext}       %
\usepackage[varvw]{newtxmath}       

\makeindex             

\usepackage{bbm}
\usepackage{url}
\usepackage{color}
\usepackage[CJKbookmarks, pdftex, bookmarksnumbered, bookmarksopen, colorlinks, citecolor=blue, linkcolor=blue,urlcolor=blue]{hyperref}

\begin{document}

\title*{Emergence of Riemannian Quantum Geometry}
\author{Hal M. Haggard, Jerzy Lewandowski, and Hanno Sahlmann}
\institute{Hal M. Haggard \at Bard College, 30 Campus Road, Annandale-On-Hudson, NY 12504, USA 
\email{hhaggard@bard.edu}, 
\and 
Jerzy Lewandowski \at Faculty of Physics, University of Warsaw
Pasteura 5, 02-093 Warsaw, Poland 
\email{jerzy.lewandowski@fuw.edu.pl}
\and
Hanno Sahlmann \at Department Physik, Friedrich-Alexander Universit\"at Erlangen-N\"urnberg (FAU), Staudtstraße 7,
91058 Erlangen, Germany 
\email{hanno.sahlmann@fau.de}
}
%
%

\maketitle

\abstract*{Each chapter should be preceded by an abstract (no more than 200 words) that summarizes the content. The abstract will appear \textit{online} at \url{www.SpringerLink.com} and be available with unrestricted access. This allows unregistered users to read the abstract as a teaser for the complete chapter.
Please use the 'starred' version of the \texttt{abstract} command for typesetting the text of the online abstracts (cf. source file of this chapter template \texttt{abstract}) and include them with the source files of your manuscript. Use the plain \texttt{abstract} command if the abstract is also to appear in the printed version of the book.}

\abstract{In this chapter we take up the quantum Riemannian geometry of a spatial slice of spacetime. While researchers are still facing the challenge of observing quantum gravity, there is a geometrical core to loop quantum gravity that does much to define the approach. This core is the quantum character of its geometrical observables: space and spacetime are built up out of Planck-scale quantum grains. The interrelations between these grains are described by spin networks, graphs whose edges capture the bounding areas of the interconnected nodes, which encode the  extent of each grain.  We explain how quantum Riemannian geometry emerges from two different approaches:  in the first half of the chapter we take the perspective of continuum geometry and explain how quantum geometry emerges from a few principles, such as the general rules of canonical quantization of field theories, a classical formulation of general relativity in which it appears embedded in the phase space of Yang-Mills theory, and general covariance. In the second half of the chapter we show that quantum geometry also emerges from the direct quantization of the finite number of degrees of freedom of the gravitational field encoded in discrete geometries. These two approaches are complimentary and are offered to assist readers with different backgrounds enter the compelling arena of quantum Riemannian geometry. }








\section{Introduction}
Since general relativity is a theory of the geometry of space and time as much as it is a theory of gravity, it seems evident that a quantum theory of geometry of some kind will be one aspect of a theory of quantum gravity. The present chapter is about the (spatial part of) the quantum geometry that arises in the quantization of gravity pursued in loop quantum gravity. 

The first half of the chapter focuses on a continuum approach and arises from a few principles with surprisingly little room for modifications. These principles are the general rules of canonical quantization of field theories, a classical formulation of general relativity in which it appears embedded in the phase space of Yang-Mills theory, and  the principle of general covariance. We will explain in  detail in this chapter how  quantum Riemannian geometry arises from this continuum approach to loop quantum gravity. 

Remarkably, quantum Riemannian geometry arises also in other contexts: many aspects of it were anticipated already by Penrose decades before the advent of loop quantum gravity, and a direct approach that  quantizes \emph{discrete} classical geometries is both illuminating and surprisingly rich. In particular, this approach looks to the finite number of degrees of freedom of the gravitational field that are captured by the geometry of Euclidean polyhedra.  We also explain how the quantization of these polyhedra gives another road to the emergence of quantum Riemannian geometry in the second half of this chapter. 

The states of quantum geometry consist of seemingly one-dimensional excitations. A basis is given by \emph{spin network states}, described by a graph, decorated with irreducible representations of SU(2) on the edges, and invariant tensors on the vertices. Remarkably they can also be read as linear combinations of quantum circuits, or as (again a linear combination of) a sort of Feynman diagram. In spin networks, the group SU(2) takes the role that Poincar\'e symmetry takes in Feynman diagrams, and the interaction vertices describe the formation of spatial volume, not the scattering and decay of particles.

One basic aspect of this geometry is the discreteness of the spectra of geometric operators, in particular that of area. A spin network edge decorated with the spin-$j$ representation contributes a quantum of area 
\begin{equation}
    a_j=8\pi \gamma \ell_P^2 \, \sqrt{j(j+1)}
\end{equation}
to any surface traversed by the edge. Here 
\begin{equation}
\ell_P^2 ={\frac{\hbar G}{c^3}}\approx 2.6 \times 10^{-70} \text{ m}^2     
\end{equation}
is the Planck area, and $\gamma$ is a parameter of the theory. This stunning scale explains the difficulty of observing quantum geometry and sets the stage for the challenge of finding its observable consequences. 

We lay out some details of the emergence of quantum Riemannian geometry in loop quantum gravity in sections \ref{sec:hf} and \ref{sec:ALrep}, and its properties in section \ref{sec:geometric_operators}. We discuss the emergence of quantum Riemannian geometry from quantizing discrete geometry in section \ref{sec:discrete_geometry}. The literature addressing both halves of this chapter is vast; rather than attempt a fully rigorous and complete account, we have opted to try to make this chapter more accessible to a researcher new to the field. We encourage readers to explore the multitude of references provided throughout the chapter for further details. 
%
\section{The holonomy-flux variables for  general relativity}
In this first section, we  present an account of the quantization of general relativity, that underlying loop quantum gravity. The result is, among other things, a quantum theory of intrinsic and extrinsic Riemannian geometry. 

The formalism is based on a phase space formulation that embeds general relativity into the phase space of SU(2) Yang-Mills theory, and an operator algebra and Hilbert space that uses no auxiliary classical structures such as a flat background metric. Therefore all the structures transform covariantly under the action of diffeomorphisms.

We will first describe the classical setup in sections \ref{sec:phase_space} and \ref{sec:hf}. Then we will come to quantization (section \ref{sec:quantization}) and finally to the resulting quantum geometry in sections \ref{sec:ALrep} and \ref{sec:geometric_operators}. More extensive accounts of the theory covered in the following sections can be found in \cite{Ashtekar:1995zh,Thiemann:2001gmi,Ashtekar:2004eh,Thiemann:2007pyv,Rovelli:2008zza}. 
\subsection{General relativity inside the phase space of Yang-Mills theory}
\label{sec:phase_space}
Consider a fixed $4$-manifold ${\cal M}$. Let   $e^I=e^0,...,e^3$ be coframes, 
and construct the metric tensor 
\begin{equation}\label{metric} g=\eta_{IJ}e^Ie^J
\end{equation}
on ${\cal M}$; here $\eta_{IJ} ={\rm diag} (-1,1,1,1)$, and is used to raise and lower capital latin indices. An SO(3,1) connection $\omega$, upon choice of a gauge, can be written as a matrix of $1$-forms  $\omega^I{}_J$, $I,J=0,...,3$, such that 
$\omega_{IJ}=-\omega_{JI}.$
The curvature $2$-form of $\omega^I{}_J$ is 
$$\Omega^I{}_J := d\omega^I{}_J + \omega^I{}_K\wedge \omega^K{}_J.$$
With these definitions, the Palatini-Holst action \cite{Holst:1995pc} is defined by 
\begin{equation}\label{PH}
S_{\rm PH}(e,\omega)=\frac{1}{4\kappa}\int_{\cal M} \epsilon_{IJKL}e^I\wedge e^J\wedge \Omega^{KL}  - \frac{1}{2\kappa\gamma}\int_{\cal M} e^I\wedge e^J\wedge \Omega_{KL}, 
\end{equation}
where $\kappa := 8\pi G$ in units with $c=1$, and $\gamma>0$ is the Barbero-Immirzi parameter, the significance of which will emerge below. 
The covariant symplectic form \cite{Lee:1990nz} derived from $S_{\rm PH}$ is
\begin{equation}\label{Symp} -\frac{1}{\kappa\gamma}\int_{\Sigma} \delta_{[1}(e^I\wedge e^J)\wedge \delta_{2]}(\omega_{IJ} - \frac{\gamma}{2}\epsilon_{IJKL}\omega^{KL}),
\end{equation}
where $\Sigma$ is any  Cauchy surface, and $\delta_1$ and $\delta_2$ are vectors tangent to the space of solutions of the resulting field equations.

A $3+1$ decomposition of ${\cal M}$ relies on a choice of a foliation  by three-dimensional surfaces. In particular, the  coframes $e^I$ should satisfy, in the dual frame $(e_0,...,e_3)$, that the vector fields $e_i=e_1,e_2,e_3$ are tangent to the leaves of the foliation. The leaves of this foliation are also assumed to be Cauchy surfaces of the corresponding  metric tensors. On each leaf there is an induced  metric 
\begin{equation}\label{q}
q=q_{ij}e^ie^j= (e^1)^2 +(e^2)^2+(e^3)^2.
\end{equation}

The tensor $q_{ij}$ lowers and raises lower-case Roman indices, which range from $1$ to $3$. 
The corresponding torsion free connection, $\Gamma^i = \Gamma^1,\Gamma^2,\Gamma^3$, is given by $1$-forms satisfying 
$$de^i + \epsilon^i{}_{jk}\Gamma^j\wedge e^k = 0,$$
where $\epsilon_{ijk}$ is the alternating symbol. While the extrinsic curvature can be expressed in terms of $1$-forms $K^i=K^1,K^2,K^3$ defined by
$$K^i_a:=e^i_b\nabla_a n^b,$$
with $n$ the normal to the leaves of the foliation, that is, $n=e_0$, with $e_0$ the timelike vector field in the frame $e_I$.   The symplectic $2$-form (\ref{Symp}) written in terms of the $3+1$ decomposition reads
\begin{equation}\label{Symp'} \frac{1}{\kappa\gamma}\int_\Sigma \left(\delta_1 E_i^a\delta_2 A^i_a - \delta_2 E_i^a\delta_1 A^i_a\right)d^3x, 
\end{equation}
where 
\begin{equation}\label{AB}E_i^a := \sqrt{{\rm\ det q}}\,e_i^a, \ \ \ A^i_a = \Gamma^i_a + \gamma K^i_a,
\end{equation}
and the frame vectors $e_i=e_1,e_2,e_3$ are  tangent to $\Sigma$ and  daul to the coframe $e^i$. The symplectic form above gives rise to the Poisson brackets
\begin{equation}
\label{eq:PB_AE}
    \{A^i_a(x),E_j^b(y)\}=\kappa\gamma \delta^j_i\delta_a^b\delta^{(3)}(x,y). 
\end{equation}

Clarification of the resulting symmetries is in order. The Palatini-Holst action (\ref{PH}) is symmetric with respect to the diffeomorphisms of ${\cal M}$ and local Lorentz rotations of the coframe field, which are accompanied by transformations of the connection $1$-forms $\omega^I{}_J$:
$${e'}^I = (g^{-1})^I_J e^J, \ \ \ \omega'^I{}_J = (g^{-1})^I{}_K \omega^K{}_L g^L{}_J + (g^{-1})^I{}_K dg^K{}_J.$$
The $3+1$ decomposition reduces that symmetry to the diffeomorphisms preserving the foliation of ${\cal M}$
and to the $SO(3)$ coframe rotations preserving the frame element $e_0$, which is orthogonal to the foliation (with respect to the metric tensor (\ref{metric})).  In handling these symmetries we make one more step, namely we consider a basis $\tau_i = \tau_1,\tau_2,\tau_3 \in \mathfrak{su}(2)$ of the Lie algebra of the group ${\rm SU}(2)$ such that 
$$[\tau_i,\tau_j]=\epsilon^k{}_{ij}\tau_k, \ \ \ \  -2{\rm Tr}(\tau_i\tau_j)=q_{ij},$$
and  use it to collect the $1$-forms  $A^i_a$ into an $
\mathfrak{su}(2)$-valued $1$-form $A$, and the vector densities $E^a_i$ into an $\mathfrak{su}(2)$-valued vector density $E$,
\begin{equation}\label{eq:AandE}
    A = A^i_a \tau_i\otimes dx^a, \ \ \ \ E = E^{ia} \tau_i\otimes \partial_i.
\end{equation}
Now, the local rotations are represented by ${\rm SU}(2)$-valued maps $g:\Sigma \rightarrow {\rm SU}(2)$, and 
\begin{equation}\label{YM} E' = g^{-1}E g, \ \ \ A' = g^{-1}A g + g^{-1}dg.   
\end{equation}

With this setup, another choice of canonical variables consistent with the limit $\gamma\rightarrow 0$ would be 
$${P}_{i}^{''a} := \kappa\sqrt{{\rm\ det q}}\, e_i^a, \ \ \ {A}^{''i}_a =  K^i_a + \gamma\Gamma^i_a .$$ 
However, the special property of the Ashtekar-Barbero variables (\ref{AB}), see \cite{Ashtekar:1986yd,BarberoG:1994eia}, is that $A^i_a$ is a connection $1$-form defined on an ${\rm SU}(2)$ bundle covering the orthonormal frame bundle over $\Sigma$.

\subsection{The classical holonomy-flux variables}
\label{sec:hf}
The local frame rotations (\ref{YM}) clearly involve non-physical degrees of freedom. What captures the geometrical (physical) degrees of freedom is the parallel transport. Consider a path $[\tau_0,\tau_1]\ni\tau \mapsto p(\tau)\in \Sigma$ and the equation
$$\frac{dh(t,t_0;p,A)}{d\tau} = - A_a(p(\tau)){\dot p}^a(\tau)h(t,t_0;p,A), \ \ \ \ h(t_0,t_0;p,A) = {\rm id} . $$
We assign to every path $p$ the corresponding parallel transport 
\begin{equation}
\label{eq:holonomy}
    h_p(A) := h(t_1,t_0;p,A) \in {\rm SU}(2),
\end{equation}
and with a slight abuse of language, we call this open-path holonomy just a holonomy. Notice, that $h_p(A)$ is independent of orientation preserving reparametrizations of $p$; meanwhile orientation reversal induces the flip, $h_{p^{-1}} = (h_p)^{-1}$, 
where $p^{-1}$ represents the same path, but with opposite orientation, and on the right we have the inverse element in ${\rm SU}(2)$. 
An important property of the holonomies is the composition rule
\begin{equation}\label{decomp}
h_{p_1\circ...\circ p_n}(A) = h_{p_1}(A)...h_{p_n}(A).     
\end{equation}
On the other hand, to every oriented $2$-surface $S\subset \Sigma$ (a submanifold) and  smearing function $f:S\rightarrow {\rm su}(2)$, we assign the flux
\begin{equation}\label{Fl}
P_{S,f}(E):= \frac{1}{2}\int E_i^af^i \epsilon_{abc}dx^b\wedge dx^c.
\end{equation}
Note that the smearing function $f$ may involve a point-dependent holonomy of $A$, chosen such that the integrand is invariant with respect to (\ref{YM}), however, one may also use smearing functions independent of $A$.

The Poisson bracket between the holonomies  and fluxes can be easily calculated and has a simple, geometrical structure. Indeed, if the path $p$ begins or ends on the surface $S$ and does not cross it anywhere else,  then, \cite{Ashtekar:1996eg},
\begin{equation}\label{thebracket}
\{h_p,P_{S,f}\}=-\kappa\gamma\frac{\sigma(S,p)}{2}
\begin{cases}
			h_pf(p(t_0)), & \text{if } p(t_0)\in S \\
            -f(p(t_1))h_p, & \text{if } p(t_1)\in S,
		 \end{cases}
\end{equation}
where
\begin{equation}
\label{eq:sigma}
    \sigma(S,p) = \begin{cases}
			1, & \text{if $p$ lies above $S$} \\
           -1, & \text{if $p$ lies below $S$}\\
           0, & \text{otherwise}.
		 \end{cases}
\end{equation}
Here the terms \emph{above} and \emph{below} in the formula should be interpreted as follows. Since $\Sigma$ and $S$ are both oriented, they divide the tangent space $T_x\Sigma$ at the intersection point $x=p\cap S$ into three sets: those tangent vectors that when appended to a positively oriented basis of $S$ give a positively oriented basis of $\Sigma$, those that give a negatively oriented one, and those that are tangent to $S$. If a positively oriented tangent to $p$ lies in the first set, we say that $p$ is above $S$, if it lies in the second, that $p$ is below $S$.

If a path $p$ can be decomposed  into paths of one of the categories considered above, then via (\ref{decomp}) and the Leibniz rule  the Poisson brackets (\ref{thebracket}) can be applied. On the other hand, the Poisson bracket (\ref{thebracket}) vanishes if the path $p$ does not intersect $S$ or if it is contained in $S$. A case that remains unaccounted for is a path $p$ intersecting the surface $S$ in infinite number of isolated points. We eliminate those by assuming a semi-analytic structure on $\Sigma$ \cite{Lewandowski:2005jk}, one that is a proper generalization of piecewise analyticity (see subsection \ref{DiffsAndDiffInv}).  

The Poisson bracket relations \eqref{thebracket} have a remarkable consequence. The variables $P_{S,f}$ do not Poisson commute \cite{Ashtekar:1998ak}. Due to the Jacobi identity we have 
\begin{equation}\label{eq:jacobi}
    \left\{ \{ {P}_{S_1,f_1}, {P}_{S_2,f_2} \}, 
    {h}_p \right\} =-  \left\{ \{ {P}_{S_2,f_2} , {h}_p \}, 
     {P}_{S_1,f_1} \right\} - \left\{ \{ {h}_p , {P}_{S_1,f_1} \}, 
      {P}_{S_2,f_2} \right\},
\end{equation}
implying a non-trivial bracket $\{ {P}_{S_1,f_1}, {P}_{S_2,f_2} \}$. Closer inspection shows that it is non-zero only for $S_1\cap S_2\neq \emptyset$ and has non-zero brackets with $h_p$ only if $S_1\cap S_2 \cap p \neq \emptyset$. Non-Poisson-commuting $P$'s are surprising at first sight since canonical commutation relations would seem to imply that the field $E$ commutes with itself. However the non-commutativity can be understood in a natural way by defining the $P$'s as (infinite dimensional) vector fields \cite{Ashtekar:1998ak} or by the observation that 
\begin{equation}
    P'_{R,f} =\frac{1}{2}\int_R d^{A} f^i \wedge E^a_i \epsilon_{abc} dx^b\wedge dx^c  
\end{equation}
for $R$ a region with boundary $\partial R=S$ differs from $P_{S,f}$ by a term that vanishes when the Gauss constraint gets taken into account \cite{Cattaneo:2016zsq}. Here $d^{A}$ is the covariant form derivative with respect to $A$. Thus $P'_{R,f}$ will reproduce the brackets \eqref{thebracket}, while explaining \eqref{eq:jacobi} due to the presence of $A$ and $E$. 

The holonomy variables allow better control of the local rotation transformations (\ref{YM}). Indeed, 
\begin{equation}\label{YMhol}
h_p(g^{-1}Ag + g^{-1}dg) =  g^{-1}(p(t_1))h_pg(p(t_0)).    
\end{equation}
In the following we will consider functions of $A$ depending solely on the holonomies along a finite number of paths in $\Sigma$. If the paths are chosen such that they form a graph $\Gamma$ (i.e., they intersect in their boundaries only) then the local rotation transformations act at the vertices of $\Gamma$ only. We will call such functions \emph{cylindrical} \cite{Ashtekar:1994mh}.

The transformation properties of the flux variables that use $A$-independent smearing functions, are 
\begin{equation}
P_{S,f}(g^{-1}Eg) = P_{S,gfg^{-1}}(E).      
\end{equation}
That is why in some cases we use holonomy dependent $f$'s, \cite{Thiemann:2000bv}, such that
\begin{equation}
P_{S,f}(g^{-1}Eg, g^{-1}Ag g^{-1}dg) = P_{S,f}(A,E).     
\end{equation}

\subsection{The quantum holonomy-flux variables} 
\label{sec:quantization}
Identifying a quantum theory corresponding to the classical structures discussed in the last section requires the choice of an algebra of kinematic quantities. The only hard requirement on the algebraic structure is that, to first order in $\hbar$, the Poisson relations \eqref{eq:PB_AE} are realized as commutators, a consistency requirement for the classical limit. This still leaves a lot of possibilities. Further reasonable requirements are diffeomorphism covariance, gauge covariance
and simplicity. 
Diffeomorphisms of $\Sigma$ and gauge transformations are generated by the constraints and act on the fields \eqref{AB}. They can hence be expected to act on the algebra underlying the quantum theory. The algebra should be closed under these transformations. The algebraic structure should also be free of fixed classical structures, such as a fixed classical metric. This is because they would not transform, they would distinguish a (gauge or coordinate) frame and thus be unnatural in a generally covariant theory.

The algebra usually chosen is the \emph{holonomy-flux algebra}, generated by elements 
\begin{equation}
 \label{HolFluxGen}   (\widehat{h}_p)^{a}{}_b, \qquad \widehat{P}_{S,f}.
\end{equation}
These can be thought of as  representing the holonomies \eqref{eq:holonomy} of $A$ along paths $p$ in $\Sigma$, 
and fluxes, \eqref{Fl}, where $f$ is a triple of smearing functions and $S$ is an oriented surface. The classical quantities transform under diffeomorphisms of $\Sigma$ in a simple way since the integrands are top forms on the submanifolds being integrated over. 
As for algebraic relations, the algebra elements inherit the relations governing the parallel transport, for example,
\begin{equation}\label{eq:star1}
    \widehat{h}_{p_2\circ p_1} =  \widehat{h}_{p_2} \widehat{h}_{p_1}, \qquad  \widehat{h}_p ( \widehat{h}_p)^\dagger =\mathbbm{1},
\end{equation}
where these equations now have to be read as matrix equations with algebra-valued entries, the dagger being matrix transpose and algebra adjoint. Similarly, the symbols $\widehat{P}_{S,f}$ inherit relations from their classical counterparts, such as 
\begin{equation}\label{eq:star2}
    \widehat{P}_{S,f_1+f_2} = \widehat{P}_{S,f_1}+\widehat{P}_{S,f_2}, \qquad \widehat{P}_{S,f}^\dagger= \widehat{P}_{S,\overline{f}}. 
\end{equation}
The key non-classical relation in the holonomy-flux algebra is the commutator 
\begin{equation}
\label{eq:commutator}
    [P_{S,f}, h_p] = 
        \begin{cases}0 & \text{ if  $ S \cap p = \emptyset $ }\\
        8\pi \ell_\text{P},\sigma(S,p)\,  h_2\, \tau_i f^i(x)\, h_1 &\text{ otherwise},  
        \end{cases}
\end{equation}
mirroring \eqref{thebracket}. 
Here $x$ is a single intersection point of $S$ and $p$, and $\sigma$ is the sign \eqref{eq:sigma} depending on the relative orientation of $p$ and $S$. If $p$ is tangential to $S$ at $x$, $\sigma$ vanishes. There is a straightforward generalization to the case of multiple intersections where the result is a sum over intersections, with each intersection a term like \eqref{eq:commutator}. Note that there is closure under \eqref{eq:commutator} since the commutator yields a sum of products of holonomy matrix elements. 
The algebra generated by the relations \eqref{eq:star1}, \eqref{eq:star2}, and \eqref{eq:commutator}, together with the Jacobi identity is called the \emph{holonomy-flux algebra} $\mathfrak{A}_{\text{HF}}$.\footnote{For a careful definition of $\mathfrak{A}_{\text{HF}}$ see, for example, \cite{Lewandowski:2005jk}. There are slightly different definitions of this algebra in the literature, depending on whether one wants to impose additional higher order commutator relations, see \cite{stottmeister2013structural,Koslowski:2011vn} for more details.}

Note that \eqref{eq:commutator} only uses ingredients such as intersection, evaluation and relative orientation that are invariant under diffeomorphisms. This implies that diffeomorphisms act as algebra homomorphisms on $\mathfrak{A}_{\text{HF}}$. 

Another important aspect of \eqref{eq:commutator} and $\mathfrak{A}_{\text{HF}}$ is that non-commutativity of spatial geometry is unavoidable. As anticipated by the classical relations \eqref{eq:jacobi}, one finds
\begin{equation}
    \left[ [ \widehat{P}_{S_1,f_1}, \widehat{P}_{S_2,f_2} ], 
    \widehat{h}_p \right] =-  \left[ [ \widehat{P}_{S_2,f_2} , \widehat{h}_p ], 
     \widehat{P}_{S_1,f_1} \right] - \left[ [ \widehat{h}_p , \widehat{P}_{S_1,f_1} ], 
      \widehat{P}_{S_2,f_2} \right] ,
\end{equation}
which implies non-trivial commutators of $\widehat{P}$'s in general. Since these objects correspond to the spatial metric geometry, one is dealing with a form of non-commutative Riemannian geometry.

Now representations of this algebra can be studied. One important representation is the \emph{Ashtekar-Lewandowski} representation \cite{Ashtekar:1994mh,Ashtekar:1994wa,Ashtekar:1995zh} to which we turn next. 
\subsection{Ashtekar-Lewandowski representation}
\label{sec:ALrep}
To obtain the Ashtekar-Lewandowski representation one can follow the traditional strategy, split the variables $(A,E)$ into ``positions" $A$ and ``momenta" $E$, and define quantum states to be functions of the positions. It will turn out that the result is unique after some assumptions.  For this purpose, we  use the algebra ${\rm Cyl}^{(\infty)}$ of cylindrical functions,  that is, the functions that can be written in the following way:
\begin{equation}\label{cyl}
\Psi(A) = \psi(h_{p_1}(A), ... , h_{p_n}(A)),    
\end{equation}   
where $p_1,...,p_n$ are arbitrary paths in $\Sigma$, the number $n$  depends on $\Psi$ and is arbitrary,  and $\psi\in C^{\infty}({\rm SU}(2)^n)$. Clearly,  these functions define a subalgebra of the algebra of all the functions of the variable $A$. Given a cylindrical function $\Psi$, the set of the paths in (\ref{cyl}) is not unique. For example, we can always subdivide a path into two, or change its orientation, or just add a new, unnecessary  path. The key observation is that two semianalytic paths can intersect one another only at a finite set of  isolated points and along a finite set of connected segments. As a consequence,  given  a cylindrical function  (\ref{cyl}), we can subdivide the paths in such a manner that the resulting paths  intersect each other at most at one or at two endpoints, that is, they describe a graph embedded in $\Sigma$. Thus, every cylindrical function  can be written in the form (\ref{cyl}) where the paths $p_1,...,p_n$ are edges of a graph $\Gamma$ embedded in $\Sigma$.  

To define the proto-Hilbert product between two states $\Psi$ and $\Psi'$, we choose a graph $\Gamma$, specified by the paths  
$\{p_1,...,p_n\}$, and such that each state can be written in the form (\ref{cyl}), then define
\begin{equation}\label{integral}
(\Psi, \Psi') := \int dg_1...dg_n \overline{\psi(g_1,...,g_n)}\psi'(g_1,...,g_n).     
\end{equation}
 Importantly, the result is independent of the choice of graph. The Hilbert space ${\cal H}_\text{AL}$ of the quantum states is defined to be the completion of ${\rm Cyl}$ in the norm defined by the  product 
$(\cdot,\cdot)$.  

Every cylindrical function $\Psi$ is promoted to a quantum operator $\widehat \Psi$ acting in ${\cal H}$ 
naturally: 
${\widehat \Psi}\Psi'=\Psi\Psi'.$
While every flux observable $P_{S,f}$ gives rise to a quantum operator ${\widehat P}_{S,f}$ defined (originally) on ${\rm Cyl}^{(\infty)}$ by
$\widehat{P}_{S,f} \Psi := i\hbar \{P_{S,f},\Psi\}.$
The properties of the ``position" operators are obvious. The flux operator is symmetric (and even essentially self-adjoint) provided the holonomies potentially involved in the definition of $f$ are all contained in the surface $S$ and the function $f$ is real valued.  

The flux operator gives rise to a quantum  spin operator $\hat{J}_i^{x[p]}$ assigned to a point $x\in\Sigma$ and a class $[p]$ of curves that begin at $x$ and share  initial segments. Given a cylindrical function $\Psi$ we write it in the form (\ref{cyl}) with a graph such that one of the edges, say $p_1$, begins at $x$ and belongs to the class $[p]$. Then, 
\begin{equation}\label{J}
\hat{J}_i^{x[p]}\Psi(A) = i\hbar\frac{d}{ds}_{|_{s=0}}\psi(h_{p_1}(A)e^{s\tau_i}, h_{p_2}(A),...,h_{p_n}(A)). 
\end{equation}
In terms of the quantum spin operators, the quantum flux operator is
\begin{align}\label{qflux}
\widehat{P}_{S,f} &= \frac{\kappa\gamma}{2}\sum_{x\in S} f^i(x)\left(\sum_{[p]{\rm\ going\ up}}\hat{J}_i^{x,p} - \sum_{[p]{\rm\ going\ down}}\hat{J}_i^{x,p}\right) \nonumber \\
&= \frac{\kappa\gamma}{2}\sum_{x\in S} f^i(x)\left(\hat{J}_i^{x,S\uparrow} - \hat{J}_i^{x,S\downarrow}  \right).  
\end{align}
The sum on $x$ ranges over the full surface $S$, and that over $[p]$ ranges over all the (classes) of edges at $x$. However, when the operator is applied to a cylindrical function, then these sums only range over the isolated intersections of the surface with the curves the function depends on, and the edges $[p]$ that overlap the curves of the cylindrical function.

 Diffeomorphism $\phi:\Sigma\rightarrow \Sigma$ naturally act on the cylindrical functions (\ref{cyl}) via
$$U_{\phi}\Psi(A) = \Psi(\phi^*A) = \psi(h_{\phi(p_1)}(A),...,h_{\phi(p_n)}(A)).$$
The operator $U_{\phi}$ is unitary in ${\cal H}$. The fluxes are also diffeomorphism covariant 
$$U_\phi \widehat{P}_{S,f}U_\phi^{-1} = \widehat{P}_{\phi^{-1}(S),\phi^*f},$$
and if $f$ involves holonomies along paths $q_1,...,q_k$, then $\phi^*f$ depends in the same way on the holonomies along $\phi^{-1}q_1,...,\phi^{-1}q_n$. It should be noted, however, that diffeomorphisms do not admit infinitesimal generators acting in the space of the cylindrical functions. 

The local rotations (\ref{YM}) also act naturally and unitarily in  the Hilbert space ${\cal H}$ via (\ref{YMhol}).  They are generated by the Lie algebra of the local rotation operators 
\begin{equation}
\int_\Sigma d^3x \widehat{\left(\left(D_a\Lambda\right) E^a\right)} = \frac{\kappa\gamma}{2}\Lambda^i(x) \sum_{x\in\Sigma}\sum_{[p]\ {\rm at}\ x} \hat{J}_i^{x,[p]} =: \frac{\kappa\gamma}{2}\Lambda^i(x) \sum_{x\in\Sigma}\hat{J}_i^{x,{\rm tot}}.      
\end{equation}
The compact notation of the final expression relates the total spin operator, just introduced, to the total spin operators $S\uparrow$, $S\downarrow$, and  $S\parallel$, which accounts for the curves contained in $S$, via
\begin{equation}
\hat{J}_i^{x,{\rm tot}} =  \hat{J}_i^{x,S\uparrow} + \hat{J}_i^{x,S\downarrow} + \hat{J}_i^{x,S\parallel}.   
\end{equation}

The algebra generated by the cylindrical functions and the flux operators, fulfills the relations of the holonomy-flux algebra $\mathfrak{A}_\text{HF}$ and, thus, forms a representation of  $\mathfrak{A}_\text{HF}$ on the Hilbert space ${\cal H}$. Moreover, spatial diffeomorphisms act in a unitary manner as described above, and leave the Ashtekar-Lewandowski vacuum, represented by the constant cylindrical function, invariant. It can be shown that this is the only representation of $\mathfrak{A}_\text{HF}$ in which the spatial diffeomorphisms are unitarily represented and that contains an invariant cyclic vector \cite{Lewandowski:2005jk,Fleischhack:2004jc}. 


If we fix a graph $\Gamma$, specified by paths $p_1,...,p_n \subset M$,  in the definition of a cylindrical function, then the Peter-Weyl theorem provides an orthonormal basis. Fix an orthonormal basis in the Hilbert space ${\cal H}_{j_I}$ of irrep $j_I$, then a basis element is defined by assigning to each edge $p_I$ the $(n^I,m_I)$-th  entry of the Wigner matrix ${D_{(j_I)}}^{m_I}{}_{m'_I}$. From this data we construct a function on ${\rm SU}(2)^n$,
\begin{equation}\label{SN0}
\psi(g_1,...,g_n) = \sqrt{(2j_1+1)...(2j_n+1)}{D_{(j_1)}}^{m_1}{}_{m'_1}(g_1)...{D_{(j_n)}}^{m_n}{}_{m'_n}(g_n).
\end{equation}
In order to control the properties of that function with respect to the gauge transformations, at each intersection point $v$ between incoming edges $p_{I_1},...,p_{I_k}$, and outgoing, say $p_{J_1},...,p_{J_l}$  we introduce tensors   ${{\iota_v}_{m'_{I_1}...m'_{I_k}}{}^{m_{J_1}...m_{J_l}}}_m$, termed intertwiners, that intertwine the corresponding representations (or the dual ones in the case of the outgoing edges) into a new irreducible representation $j_v$,\footnote{Let $D_\otimes=D_{(I_1)}\otimes\ldots\otimes D_{(I_k)}\otimes D^*_{(J_1)}\otimes\ldots\otimes D^*_{(J_l)}$ be the tensor product representation. The intertwiner ${{\iota_v}_{m'_{I_1}...m'_{I_k}}{}^{m_{J_1}...m_{J_l}}}_m$ is given by the matrix elements of an equivariant map  $$
    \iota: \mathcal{H}_\otimes \longrightarrow \mathcal{H}_{j_v}, \qquad D_{(j_v)}\circ \iota =\iota \circ D_{\otimes}.
$$ The final $m$ index of $\iota$ labels the states of $\mathcal{H}_{j_v}$; below we will also consider gauge-invariant states with $j_v=0$ and this index will not always appear. 
}   and contract them correspondingly,
\begin{equation}\label{SN}
\otimes_{v} \iota_{v_\alpha} \lrcorner \, \otimes_I\, \sqrt{2j_I+1}D_{(j_I)}.     
\end{equation}
In particular, if we take intertwiners into the trivial representation only, the ${\rm SU}(2)$ invariants, then we obtain a gauge invariant cylindrical function, also referred to as a \emph{spin network} \cite{Baez:1994hx}.


In fact, since any operator in one of the families $\{\hat{J}_i^{y,[p]}\hat{J}^i{}^{y,[p]}\}_{y,[p]}$ and $\{\hat{J}^{y,\rm tot}\}_{y}$  commutes with any other one in these families, there exists a joint eigenbasis for all of them. This basis consists of the states \eqref{SN}. It follows, in particular, that 
an operator ${\widehat {\cal O}}$ satisfying 
\begin{equation}\label{comutation}
[{\hat {\cal O}}, \hat{J}_i^{y,[p]}\hat{J}^i{}^{y,[p]}]=0=[{\hat {\cal O}}, \hat{J}^{y,\rm tot}] \qquad \text{ for all } y,[p],     
\end{equation}
must be diagonal in this basis as well. 

\subsection{Diffeomorphisms and diffeomorphism invariance}
\label{DiffsAndDiffInv}
 In LQG we distinguish between many types of diffeomorphisms and  smoothness. In this subsection we discuss  the semi-analytic category and the corresponding diffeomorphisms with respect to  which the quantum theory presented above is invariant. Later, in section \ref{sec:other_phases}, we mention other directions of theory construction based on other diffeomorphisms, which have not been fully exploited, although they have provided interesting and sometimes extremely mathematically sophisticated results. 

In one dimension, a semi-analytic function $f:\mathbb{R}\rightarrow \mathbb{R}$ is any differentiable function that is piece-wise analytic. To generalize this definition to higher dimensions a suitable notion of ``piece-wise" is required.  The idea  invented for the rigorous formulation of LQG diffeomorphism invariance was to introduce a new category of differentiability through the theory of semi-analytic sets and  semi-analytic partitions \cite{ASNSP_1964_3_18_4_449_0,PMIHES_1988__67__5_0,Lewandowski:2005jk}. If a function defined on $\mathbb{R}^n$ is analytic when restricted to every element of some semi-analytic partition of $\mathbb{R}^n$, then we  call it semi-analytic. The local version of this property provides the general definition \cite{Lewandowski:2005jk}. Notice that every semi-analytic partition of $\mathbb{R}^n$ contains open subsets as well as subsets of lower dimensions, for example: cube interiors, face interiors, site interiors and vertices. The family of real-valued semi-analytic functions defined in ${\mathbb{R}^n}$ has all the properties needed to define semi-analytic manifolds and submanifolds, as well as semi-analytic diffeomorphisms \cite{Lewandowski:2005jk}. The semi-analytic category  combines important  properties  of the analytic category on the one hand, and differentiable (or smooth) category, on the other. In the semi-analytic category, every finite family  of  curves contained in a  manifold $M$ can be cut into finitely many pieces that fix an embedded graph in $M$.  Given a $2$-surface $S$ in a $3$-manifold, every curve can be cut into finitely many pieces such that each of the pieces either does not intersect $S$, or intersects at a single point, or overlaps $S$. These properties are adopted from the analytic category. On the other hand, local (semi-analytic) diffeomorphisms  still exist for arbitrarily small compact supports, as in the case of the differentiable category.      

The quantum holonomy and quantum flux operators defined above are diffeomorphism covariant, that is, they are unitarily mapped by diffeomorphisms into other holonomy and flux operators. In contrast, the total quantum volume operator, or the integral of the quantum scalar curvature operator, are diffeomorphism invariant. What are other diffeomorphism invariant operators?  It turns out that  every self-adjoint diffeomorphism invariant operator that contains all the cylindrical functions in its domain preserves the graphs: a cylindrical function defined by using a given graph is mapped into a cylindrical function that can be defined with the same graph \cite{Ashtekar:1995zh}. This theorem underlies  `algebraic LQG'
\cite{Giesel:2006uj,Giesel:2006uk,Giesel:2006um,Giesel:2007wn}.       

The diffeomorphically invariant characterization of an immersed graph includes its global and local properties. The former are the generalized knots and links formed by graph edges. The latter describe how edges meet at vertices. Every generic triple of edges meeting at a vertex  can be diffeomorphically transformed into intersection of the  three axes of any fixed coordinate system. If there is a fourth edge, the remaining freedom to scale the axis can be used to fix  the position of the edge arbitrarily inside a certain region representing one-eighth of the angle range. Then, a location of a fifth edge meeting at this vertex (if it is there) becomes a diffeomorphism invariant feature. In this manner, starting from valency five, graph vertices contribute a continuous set of diffeomorphism invariant degrees of freedom, see \cite{Grot:1996kj}.

\section{Quantum geometry}
\label{sec:geometric_operators}
The states of the Ashtekar-Lewandowski representation are states of extrinsic and intrinsic geometry of the spatial slice $\Sigma$. This can be seen most directly when probing them with operators representing various aspects of this geometry. In the present chapter, we will restrict ourselves to the intrinsic Riemannian geometry of the spatial slice and consider the corresponding operators and their properties in sections \ref{sec:area} to  \ref{sec:regge}. What emerges is a geometry that associates the vertices of spin networks with the volume of spatial regions and the edges with areas.

Somewhat surprisingly, the quantum geometry obtained through the study of general relativity can also be seen as the quantization of piecewise flat discrete geometries. In section  \ref{sec:polyhedral_states}, we briefly explain how this picture arises in loop quantum gravity. This also lays groundwork for section \ref{sec:discrete_geometry} of this chapter, where quantum discrete geometries will be developed starting from their classical foundations.  

In section \ref{sec:quantum_reduced}, we discuss a certain limit of the quantum geometric states and operators that simplifies calculations and can be used in quantum cosmology. Finally, section \ref{sec:other_phases} briefly presents some extensions of the picture of quantum geometry in the presence of matter, in other representations of holonomies and fluxes, and in some extensions of the entire formalism. 
\subsection{The area operators}
\label{sec:area}
The first geometric operator to be defined in loop quantum gravity, and perhaps the simplest, is the area operator \cite{Rovelli:1994ge,Ashtekar:1996eg}. It is very natural in this setting because the area 2-form is essentially the length (in internal space) of the field $E$ \cite{Rovelli:1993vu}.   

Consider a $2$-dimensional surface $S\subset \Sigma$. For each classical  frame density field $E_i^a$, cf. (\ref{AB}), the area element  $d^2s$ in $S$ can be approximated by the fluxes of (\ref{Fl}) through  small pieces $S_n$ that set a partition of $S$. Indeed, for every function $F$ defined on $S$, 
\begin{equation}
 \int_S F d^2s = \lim_{N\to\infty} \kappa\gamma \sum_{n=1}^N F(x_n)\sqrt{P_{S_n,\tau_i}P_{S_n,\tau_i}q^{ij}},\ \ \ \ \ x_n\in S_n,  
\end{equation}
provided in the limit the partition gets uniformly finer.  In the quantum theory we replace the classical fluxes by the quantum flux operators and derive the following integral in the quantum theory, \cite{Ashtekar:1996eg},
\begin{equation}
\int_S F \widehat{d^2s} = \frac{\kappa\gamma}{2}\sum_{x\in S}F(x)  \sqrt{-\Delta_{S,x}}\ ,
\end{equation}
where
\begin{equation}
-\Delta_{S,x} := q^{ij}\left(\hat{J}_i^{xS\uparrow}-\hat{J}_i^{xS\downarrow}\right)\left(\hat{J}_j^{x,S\uparrow}-\hat{J}_j^{x,S\downarrow}\right) .
\end{equation}
As in the case of  the flux operator, $x$ ranges over the points of $S$, however,   when the operator is applied to a cylindrical function, then only isolated intersections of the surface with the curves that the function depends on contribute. Remarkably, the operator is well defined: no infinities appear.

The eigenstates and eigenvalues of the operators $-\Delta_{S,x}$ follow from the algebraic properties of the spin operators, they are 
\begin{equation}\label{-Delta}
\lambda = \hbar^2\left(2j^\uparrow(j^\uparrow+1) + 2j^\downarrow(j^\downarrow+1) - j^{\downarrow\uparrow}(j^{\downarrow\uparrow}+1)\right),
\end{equation}
where
$j^\uparrow, j^\downarrow = 0, \frac{1}{2}, ..., \quad \text{ and } \quad
j^{\downarrow\uparrow} = |j^\uparrow - j^\downarrow|, |j^\uparrow - j^\downarrow|+1, ..., |j^\uparrow + j^\downarrow|,$
and we have used the same notation as in the definition of the flux operator, Eq.~(\ref{qflux}). 

The eigenvalues of the area operator, 
$$\hat{A}_S = \int_S \hat{d^2 s} = \frac{\kappa\gamma}{2}\sum_{x\in S}F(x)  \sqrt{-\Delta_{S,x}},    $$
are all finite sums of the  elementary areas
$
a_S = \frac{\kappa\gamma}{2}\sum_v \sqrt{\lambda_v} , 
$ 
where each $\lambda_v$ is of the form (\ref{-Delta}), \cite{Ashtekar:1996eg}.     

A generic case of the eigenvalue (\ref{-Delta}) is at the intersection of a surface $S$ with a transversal curve that is cut at the intersection point and divided into two edges, both oriented to be outgoing. Then $j^\uparrow=j^\downarrow=j$ and $j^{\downarrow\uparrow}=0$, and the corresponding eigenvalue of the area is $a_s = \kappa\gamma\hbar \sqrt{j(j+1)}.$
The simplest case, on the other hand, is a single outgoing edge, that amounts to $j^\uparrow=j=j^{\downarrow\uparrow}$,  $j^\downarrow=0$ and
$a_S = \frac{1}{2}\kappa\gamma\hbar\sqrt{j(j+1)}.$

If $S$ is a connected closed surface and splits $\Sigma$ into two disconnected parts and if we consider only gauge invariant cylindrical functions, then  there are additional constraints, namely
$$ \sum_v j_v^\uparrow \in \mathbb{N}, \ \ \text{ and } \ \ \sum_v j_v^\downarrow \in \mathbb{N} .$$

The quantum area operator is  invariant under local rotations, that is, $[\hat{A}_S,\hat{J}_i^{x,\ \rm{tot}}] =0$,
and it is covariant with respect to diffeomorphisms: $U_\phi \hat{A}_S U^{(-1)}_\phi = \hat{A}_{\phi^{-1}A}.$
The operator $\hat{q}_x$ also commutes with it,
$[\hat{A}_S, \hat{J}_i^{y,[p]}\hat{J}^i{}^{y,[p]}]=0.$

In some LQG models with boundary, each  $2$-surface is equipped with a distinguished $\mathfrak{su}(2)$-valued function $r:S\rightarrow {\rm su}(2)$, and only the $E^a_i$ frames are allowed, such that the vector field  
$n^a:= {r^iE^a_i}/{\sqrt{{\rm det}\, E}}$
is normal to $S$. Then, we  define another area operator, \cite{FernandoBarbero:2009ai},
$$\hat{A'}_S := \int_S |r^i E_i^a \epsilon_{abc}dx^b\wedge dx^c| = \frac{\kappa\gamma}{2}\sum_x|\hat{J}_i^{x,S\uparrow}r^i + \hat{J}_i^{x,S\downarrow}r^i|. $$
The advantage of this operator is its equidistant spectrum: the eigenvalues are
$0$ and $\frac{\kappa\gamma}{2}N$, with $N\in \mathbb{N}.$ This is compatible with  quantum entropy calculations on the 2-surface (for generic intersections of the curves with the surfaces, $N$ is even).  

\subsection{The angle operator}
\label{sec:volume}
Having at our disposal the quantum triad-flux operator and the quantum area operator, we can define a quantum  operator for the scalar product between the unit vectors normal to surfaces. 

Consider two oriented surfaces $S, S'\subset \Sigma$ and a point $x\in S\cap S'$.  We will start with constructing  suitable classical expressions for unit normals  at $x$. Let $S(\epsilon)\subset S$ be a disc  of  coordinate radius $\epsilon$  centered at $x$. The limit
\begin{equation}
n_{Sx}{}_i := \lim_{\epsilon\to 0}{P_{S(\epsilon),\tau_i}}/{A_{S(\epsilon)}}     
\end{equation}
provides the components of the unit co-vector $n_{Sx}{}_a=n_{Sx}{}_ie^i_a$ orthogonal to $S$ and $x$. The scalar product with the other co-normal $n_{S'x}{}_ie^i$ does not involve $e_i$ anymore, because $q_{ij}= \delta_{ij}$, and 
$
n_{Sx}{}_a n_{S'x}{}_b q^{ab} = n_{Sx}{}_i n_{S'x}{}_j  q^{ij} = 1.    
$

Now, turn to the quantum geometry, and consider the operator  
$
{\hat{P}_{S(\epsilon),\tau_i}}/{\hat{A}_{S(\epsilon)}}.$
The numerator and denominator commute,      
$
[\hat{P}_{S(\epsilon),\tau_i},\ \hat{A}_{S(\epsilon)}] =  0  
$,
which makes the quotient independent of the ordering.  The limit 
\begin{equation}
\hat{n}_{Sx}{}_i = \lim_{\epsilon\to 0}\frac{\hat{P}_{S(\epsilon),\tau_i}}{\hat{A}_{S(\epsilon)}} =
\frac{\hat{J}_i^{x_0S\uparrow}-\hat{J}_i^{x_0S\downarrow}}{\sqrt{q^{ij}\left(\hat{J}_i^{x_0S\uparrow}-\hat{J}_i^{x_0S\downarrow}\right)\left(\hat{J}_j^{x_0,S\uparrow}-\hat{J}_j^{x_),S\downarrow}\right)}}
    \end{equation}
is well defined  on the states spanned by the eigenstates  of the operator $\Delta_{S,x}$ of non-zero eigenvalues, and  satisfies, now on the quantum level,  $
q^{ij}\hat{n}_{Sx}{}_i \hat{n}_{Sx}{}_j q^{ij} = 1.$
Finally, an operator giving the cosine of the angle between the surfaces $S$ and $S'$ at the point $x$  may be defined by
\begin{equation}
\cos\left({\hat{\alpha}_{SS'x}}\right) := \frac{1}{2}q^{ij}\left(\hat{n}_{Sx}{}_i \hat{n}_{S'x}{}_j + \hat{n}_{S'x}{}_i \hat{n}_{Sx}{}_j \right).    
\end{equation}
For this operator to give a well defined state,  after the action on a spin-network state, it is necessary and sufficient that the graph has a non-trivial and generic  intersection with each of the surfaces $S$ and $S'$. Indeed, then $0$ is not in the spectra of the area operators $A_S$ and $A_{S'}$.  This definition is inspired by \cite{SethMajor}, however, it has been reformulated to work in the diffeomorphism covariant framework of the cylindrical function states, with the consequence that it has somewhat different properties.   

\subsection{The volume operators}
\label{sec:QuantumVol}
The $3$-volume element defined  by the variables $(A,E)$ is $d^3 x \sqrt{|{\rm det} E|}$. The integral of an arbitrary function $F$ defined on $\Sigma$ can be approximated by using the fluxes through surfaces defined by a partition of $\Sigma$, with $\{x_n\} =  S^a_n\cap S^b_n \cap S^c_n$, we have 
\begin{equation}
\int_\Sigma d^3 x \sqrt{|{\det} E|} F  = \lim_{N\to\infty} \sum_n F(x_n) \sqrt{|\frac{1}{3!}\epsilon^{ijk}\epsilon_{abc}P_{S^a_n,\tau_i}P_{S^b_{n},\tau_j}P_{S^c_n,\tau_k}|}.
\end{equation}
Remarkably, the simple replacement of the classical fluxes by quantum operators again provides a well-defined finite operator in the Hilbert space ${\cal H}$ in the domain of the cylindrical functions, \cite{Rovelli:1994ge,Ashtekar:1994wa,Lewandowski:1996gk,Ashtekar:1997fb}.  However, in this case, the result strongly depends on the choice of surfaces, in particular it breaks the diffeomorphism invariance. Two different choices are pursued in the literature. They lead to two different  diffeomorphism invariant quantum volume operators. In both cases the regularizing $2$-surfaces are adjusted to a given graph, used  in the construction of a cylindrical function. In the first case, the \emph{internal regularization} \cite{Lewandowski:1996gk,Ashtekar:1997fb}, the triples of $2$-surfaces intersect in the vertices of the graph (including spurious vertices in a refined graph). In the second case, the \emph{external regularization} \cite{Rovelli:1994ge},  the $2$-surfaces form cells that contain the vertices inside. We present both results below.      

In both regularizations the quantum volume operator has a similar form with 
\begin{equation}\label{volume}
\sqrt{|\widehat{{\det} E|}}(x) = (a_0\kappa\gamma)^{\frac{3}{2}}\sum_{y\in\Sigma}\delta^{(3)}(x,y)\sqrt{\hat{q}_y}\ , 
\end{equation}
here  $x$ runs through the entirety of $\Sigma$. The two regularizations differ in how they treat and define $\hat{q}_x$. 
Let us begin with the internal regularization. The diffeomorphism invariance can be ensured by a suitable averaging with respect to some family of choices and the resulting $\hat{q}_x$ is given by
\begin{equation}
\hat{q}_x^{\text{int}} = \left|
\frac{1}{3!}\sum_{e,e',e"}\epsilon^{ijk}\epsilon(e,e',e")\hat{J}^{x,e}_i\hat{J}^{x,e'}_j\hat{J}^{x,e"}_k\right|,
\end{equation}
with $\epsilon(e,e',e")=\pm 1$ or $0$ depending on the orientation of the tangent vectors at the point $x$, and the sum is over all triples of curves starting at $x$. Note that  for cylindrical functions, only $x$, which is one of the  vertices of a corresponding graph, and $e,e',e"$ that overlap edges of the graph at $x$ contribute. The orientation-sensitive factor $\epsilon(\cdot,\cdot,\cdot)$ annihilates all the planar triples, that is, planar vertices. In non-degenerate cases it gives signs to the corresponding terms. The constant $a_0$ appearing in \eqref{volume} is arbitrary and depends on the measure used for the averaging.  

For the external regularization, one instead obtains   
\begin{equation}
\hat{q}_x^{\text{ext}} = 
\frac{1}{3!}\sum_{e\not=e'\not=e"\not=e}|\epsilon^{ijk}\hat{J}^{x,e}_i\hat{J}^{x,e'}_j\hat{J}^{x,e"}_k|,
\end{equation}
with the constant $a_0$ again arbitrary. 

Each of the operators  $\hat{q}_x^{\text{int}}$ and $\hat{q}_x^{\text{ext}}$ satisfies the commutation relations (\ref{comutation}), hence they preserve the corresponding spin-network subspaces characterised above.  
And each of the operators is diffeomorphism covariant, in the sense that 
$$ U_\phi\hat{q}_x U_\phi^{-1} = \hat{q}_{\phi^{-1}(x)}, \ \ \ \ U_\phi\hat{q}'_x U_\phi^{-1} = \hat{q}'_{\phi^{-1}(x)}. $$
There is a close  relationship between the volume operator \eqref{volume}, with either regularization, and the quantization of volume based on discrete geometry \cite{Bianchi:2011ub}, which we will turn to in section \ref{sec:quantum_tet}. In a nutshell, the volume spectrum one obtains from quantizing the phase space of Euclidean tetrahedra is very close to that of the operator \eqref{volume} when acting on a 4-valent vertex. This convergence of different approaches is surprising and reassuring.  

While these volume operators look similar, there is an essential qualitative difference between them. The operator $\hat{q}_x^{\text{ext}}$ depends  on a number of different edges meeting at a vertex, however it is insensitive to the relationships between their tangent vectors: all the edges may be tangent to each other, or contained in a single tangent plane.
The operator $\hat{q}_x^{\text{int}}$ is useful in approaches that assume the existence of an underlying manifold, and provide it with quantum geometry. Indeed, we expect the frame field determinant to distinguish planar from generic triples of vectors.\footnote{In fact, the presence of the factor $\epsilon(\cdot,\cdot,\cdot)$ complicates the analysis of $\hat{q}_x$ considerably, as it affects the spectrum and for large valence many 
sign configurations arise. For a 7-valent vertex, there are already $\mathcal{O}(10^6)$ \cite{Brunnemann:2007ca,Brunnemann:2010yv}.}
The second operator,  $\hat{q}_x^{\text{ext}}$, on the other hand, is used in approaches where the manifold emerges together with geometry from a yet more fundamental discrete structure.   

The spectrum of the operator \eqref{volume} is not known in general, but there are extensive partial analytic \cite{Thiemann:1996au,Brunnemann:2004xi,Giesel:2005bk,Giesel:2005bm,Brunnemann:2010yv,Schliemann:2013oka,Bianchi:2011ub,BianchiHaggardVolLong} and numerical \cite{Brunnemann:2007ca} results. We will touch on some of these below. 

Consistency checks on the quantization of volume have been proposed based on classical equalities involving the volume and fluxes \eqref{Fl}, \cite{Giesel:2005bk,Giesel:2005bm}. They concern both the averaging constant $a_0$ and the regularization scheme.   
In the case of the internal regularization, the results on the constant $a_0$ were ambiguous, namely $a_0 = \frac{1}{2}$ or $a_0= 1/2 (2)^\frac{1}{3}$. The external regularization case was disfavored in some cases.     

We will consider the action of the operators in some special cases. The simplest non-trivial case is a $3$-valent vertex, generic (i.e. non-planar) for  $\hat{q}_x^{\text{int}}$ and arbitrary for the $\hat{q}_x^{\text{ext}}$. In that situation both regularizations give rise to the same operator:
\begin{equation}\label{JJJ}
\epsilon^{ijk}\hat{J}^{e}_i\otimes\hat{J}^{e^\prime}_j\otimes\hat{J}^{e^{\prime \prime}}_k : \ \left(V_e\otimes V_{e'}\otimes V_{e^{\prime \prime}}\right)^{j}_m\ \rightarrow\ \left(V_e\otimes V_{e'}\otimes V_{e^{\prime \prime}}\right)^{j}_m
\end{equation}
defined in each subspace of a triple of irreducible SU$(2)$ representations $V_e\otimes V_{e'}\otimes V_{e^{\prime \prime}}$ characterised by a total spin $j$ and an eigenvalue $m$ of $\hat{J}_3^{e}+ \hat{J}_3^{e'} + \hat{J}_3^{e^{\prime \prime}}$.    
In the case $j=0$
the operator vanishes. That came as a surprise to those who discovered the volume operator of loop quantum gravity, who had wanted to restrict the theory to the $3$-valent spin-network states (famous due to Penrose \cite{Penrose1971,Penrose1972}), but in retrospect it is clear that these states correspond to degenerate, planar geoemetries, cf. section \ref{sec:discrete_geometry}.  (In fact, the result holds for an arbitrary $n$-valent vertex if the space of the invariant intertwiners, that is, the space of states with $j=0$, is $1$-dimensional \cite{Thiemann:1996au}.)

The spectrum for the three-valent case for non-gauge-invariant states is not known in closed form. But, since it is important, we outline a few special cases here. 
The operator   
$\hat{q}=|\epsilon^{ijk}\hat{J}_i\otimes\hat{J}_j\otimes\hat{J}_k|,$
 acts on the tensor product space $V_{j}\otimes V_{j'}\otimes V_{j^{\prime \prime}}$ and  
$$ {\rm Tr}\left(\hat{q}^2\right) = \frac{2}{9}j(j+1)j'(j'+1)j^{\prime \prime}(j^{\prime \prime}+1)(2j+1)(2j'+1)(2j^{\prime \prime}+1). $$
This gives an average eigenvalue squared    
$ \langle\hat{q}^2\rangle = \frac{2}{9}j(j+1)j'(j'+1)j^{\prime \prime}(j^{\prime \prime}+1). $
Special cases defined by $j_*=j_1+j_2+j_3-1$ give the eigenvalues $q$ of $\hat{q}$ as 
\begin{align}\label{eq:spectrum_threevertex}
    j= j_* \quad &\implies \quad  q = \pm\sqrt{j_1j_2j_3(j_*+1)},\\
    j= j_*-1 &\implies   q = 0, \pm\sqrt{j_*(4j_1j_2j_3 -j_1j_2 - j_2j_3 -j_3j_1) + j_1j_2j_3}.
\end{align}

Another important case is defined by a gauge invariant $4$-valent vertex $v$. In this case $\hat{q}$ acts on 
${\rm Inv}\left(V_j\otimes V_{j'}\otimes V_{j"}\otimes V_{j'''}\right)$, the subspace of $\left(V_j\otimes V_{j'}\otimes V_{j"}\otimes V_{j'''}\right)$ invariant under the diagonal action of SU$(2)$. In this case
\begin{equation}
   \hat{q}_v^{\text{ext}} = 4|\epsilon^{ijk}\hat{J}^{e}_i\otimes\hat{J}^{e'}_j\otimes\hat{J}^{e"}_k| \quad \text{and} \quad \hat{q}_v^{\text{int}} = \chi(e,e',e",e''')|\epsilon^{ijk}\hat{J}^{e}_i\otimes\hat{J}^{e'}_j\otimes\hat{J}^{e"}_k|,
\end{equation}
where $\chi$ depends on a diffeomorphism class of the vertex, and takes values $\chi(e,e',e",e''') \in \{0,1,2,3,4\}.$
For this case,  the spectrum is non-degenerate, and there is a volume gap, i.e., a smallest non-zero volume eigenvalue \cite{Brunnemann:2007ca,BianchiDonaSpeziale,BianchiHaggardVolLong}.  The scaling of the smallest non-zero and largest volume eigenvalues has been derived in \cite{BianchiHaggardVolLong} and nicely corresponds with the classical geometry of tetrahedra. For much more discussion of these results see section \ref{sec:discrete_geometry} and \cite{Bianchi:2011ub,BianchiHaggardVolLong}.

For vertices of valence higher than 4, there are only numerical results for the volume spectrum. In \cite{Brunnemann:2007ca}  spectra of \eqref{volume} for valences 5 to 7, and spins up to $j_\text{max}=13/2$ are calculated. Particular attention is paid to the lowest non-zero and the highest eigenvalue. It is found that the scaling behavior of the smallest non-zero eigenvalue with $j_\text{max}$ depends crucially on the sign configuration coming from the embedding of the vertex and the factor $\epsilon(\cdot,\cdot,\cdot)$. Generically, the volume gap increases with increase of $j_\text{max}$. But there are sign configurations in which the volume gap shows the opposite behavior: in these cases the lowest non-zero eigenvalue stays constant or decreases with increasing maximal spin. In fact, there are indications that the spectral density near zero may increase exponentially for odd valence, but the numerical data is not conclusive. 
The maximal eigenvalue increases with $j_\text{max}$ for all sign configurations, but the rate depends on the sign configuration. 
\subsection{The inverse metric tensor operator}
The inverse metric tensor in the connection frame variables is 
$q^{ab}=q^{ij}e_i^ae_j^b = {E_i^aE_j^b}/{|{\rm det}\, E|}.$
To simplify $q^{ab}$ the idea  is to probe it with a $1$-form $\omega$, construct a density of  weight $1$ that does not involve the inverse of $E^a_i$, and smear. The result is the observable 
$$\int_\Sigma d^3x \sqrt{{\rm det}\, q}\sqrt{q^{ab}\omega_a\omega_b} = \int_\Sigma d^3x \sqrt{E^{a}E^{b}\omega_a\omega_b}. $$
The corresponding operator
$$\int_\Sigma d^3x \sqrt{\hat{E}^{a}\hat{E}^{b}\omega_a\omega_b}, \ \ \ \ \ \hat{E}^a_i = -i\kappa\gamma\hbar {\delta}/{\delta A^i_a}, $$
applied to spin-network functions (\ref{cyl}) and (\ref{SN0}) turns out to be well defined and the result is clear \cite{Ma:2000au}---it acts as multiplication by the eigenvalue
\begin{equation}
\kappa\gamma\hbar \sum_{I=1}^n \sqrt{j_I(j_I+1)} \int_{p_I}|\omega_a \dot{p}^a_I|dt.    
\end{equation}
This operator was used in the quantization of the Hamiltonian of the Klein-Gordon field \cite{Lewandowski:2015xqa}, it may be applied in the case of vector fields too.  
\subsection{The length operator} 
\label{sec:length}
The orthogonal coframe  $e^i_a$ is expressed by the  densitised frame $E^a_i$ in a somewhat complicated way,
\begin{equation}
e^i_a = \frac{1}{2}\epsilon_{abc}\epsilon^{ijk}{E^b_jE^c_k}/{\sqrt{|{\rm det}|\,E}},
\end{equation}
which appears to complicate quantization. 
However, the discouraging denominator can be absorbed by the  Poisson bracket with the volume observable \cite{Thiemann:1996aw,Bianchi2009},
\begin{equation}
e^i_a = \frac{2}{\kappa\gamma}\{A^i_a, {V}\}, \ \ \ \ \ {V}:= \int_\Sigma d^3x \sqrt{|{\rm det}\,E|}.  
\end{equation}
Moreover, the length of a curve $p$ can be expressed in terms of the holonomies, the volume, and the Poisson bracket, \cite{Thiemann:1996at},
\begin{equation}
\int_p \sqrt{e^ie^j q_{ij}}  =  \kappa\gamma\lim_{N\to \infty}\sum_{n=1}^N \sqrt{-2{\rm Tr}\left(h^{-1}_{\Delta p_n} \{h_{\Delta p_n}, {V}\}h^{-1}_{\Delta p_n} \{h_{\Delta p_n}, {V}\}\right)}\ ,  
\end{equation}
where 
$p=\Delta p_1\circ ... \circ \Delta p_n$
is a partition uniformly refined as $n\to \infty$. 
After quantization, the quantum length operator becomes (one of the possibilities, \cite{Thiemann:1996at}) 
\begin{equation}
\int_p \widehat{\sqrt{e^ie^j q_{ij}}}  =  \frac{\kappa\gamma}{\hbar}\sum_{x\in \Sigma} \sqrt{-2{\rm Tr}\left(\left(h^{-1}_{\Delta p_x} [h_{\Delta p_x}, \hat{V}]\right)^\dagger \left(h^{-1}_{\Delta p_x} [h_{\Delta p_x}, \hat{V}]\right)\right)}.
\end{equation}
When this operator is applied to a spin-network state, only a term corresponding to $x$ at a vertex of the spin network  contributes; this follows from the properties of the quantum volume operator. 

\subsection{The Ricci scalar operator \`a la Regge}
\label{sec:regge}
The Riemann curvature tensor is defined by second derivatives of the triad $E^a_i$. There are currently no proposals for the quantization of such expressions that are  diffeomorpism invariant on the space of cylindrical functions or the spin-network states. Remarkably, however, the integral of the Ricci scalar can be expressed as the  limit of an expression that only uses the lengths of curves (hinges), and the deficit  angles between surfaces \cite{Regge1961}. These observables are available, as described in the previous sections.  Specifically,  consider a triangulation of $\Sigma$. It consists of three-dimensional tetrahedral cells, two-dimensional triangular faces, one-dimensional edges, and zero-dimensional vertices. Given a metric tensor $q$, we assign to every edge $e$ in the triangulation, the following data:(1.) the length $l_e$; and (2.) the total angle $\theta_e$ given by the sum of the angles between the faces of the  simplices that contain $e$, at a point $x\in e$.  
The following number converges to the integral of the scalar curvature of $q_{ab}$, \cite{Regge1961}, when we refine the triangulation 
\begin{equation}
\sum_e l_e\left(2\pi - \theta_e\right)\ \rightarrow\ \int_\Sigma d^3 \sqrt{{\rm det}\, q} R.   
\end{equation}
As we know from the previous sections the operator 
$
\sum_e 2\pi\hat{l}_e  - \frac{1}{2}\hat{l}_e\hat{\theta}_e - 
\frac{1}{2}\hat{\theta}_e \hat{l}_e 
$
is well defined in ${\cal H}$ and self-adjoint, \cite{Alesci:2014aza}. It may have a well defined limit in some weak sense; the problem is its dependence on the original triangulation, and series of refinements. These difficulties can be removed by an extra recipe of averaging with respect to the choices made \cite{Alesci:2014aza}.    

\subsection{Polyhedral states}       
\label{sec:polyhedral_states}
The picture of quantum states of geometry that emerges from loop gravity is that of curves forming a graph that contribute the flux of surface area and intersect in vertices from which  quantized volumes arise. Even more specific geometric structure is brought by spin-network states constructed from so-called \emph{coherent intertwiners} \cite{Livine:2007vk}. These intertwiners are defined as follows. Consider a vertex $v$ of a graph whose edges are labelled by spins. Orient the edges to be outgoing and number them 
$I=1,...,n_v$. For every edge $p_{I}$ choose a normalized vector $u_{I}\in \mathfrak{su}(2)$  and a vector  $|j_{I},u_{I}\rangle$  in  the corresponding representation $\mathcal{H}_{j_{I}}$ (also normalised) such that  
\begin{equation}
u^i_I\hat{J}_i^{v,[p_I]}\, {D_{(j_I)}}^{m}{}_{m'}(h_{p_I})|j_I,u_I\rangle^{m'} = \hbar j_I\, {D_{(j_I)}}^{m}{}_{m'}(h_{p_I})|j_I,u_I\rangle^{m'},   
\end{equation}
and the closure condition 
\begin{equation}
\sum_{I=1}^{n_v} j_{I} u_{I} = 0,     
\end{equation}
is satisfied. The corresponding coherent intertwiner $\iota(j_{1},u_{1};...;j_{I_{n_v}},u_{I_{n_v}})^{m_1...m_{n_v}}$ is given by the average of $\otimes_{I=1}^{n_v}|j_I,u_I\rangle$ with respect to  rotation  action  of SU$(2)$. There exists a resolution of the identity in ${\rm Inv}\left(\otimes_{I=1}^{n_v}V_{I}\right)$ by the coherent intertwiners
\begin{equation*}
{1}_{m'_1...m'_{n_v}}^{m_1...m_{n_v}} = \int d\mu(u_1,...,u_{n_v})   
 \iota^{m_1...m_{n_v}}
 \iota^\dagger_{m_1'...m_{n_v}'}.
\end{equation*}
A geometric interpretation was provided in  \cite{BianchiDonaSpeziale}, where they found that the Minkowski theorem ensures existence of a polyhedron in $\mathbb{R}^3$ with facets orthogonal to the vectors $u_I$ and facet areas $j_I$, provided the closure condition is satisfied. In this manner a quantum polyhedron can be associated to each vertex of a spin-network.  
The polyhedra assigned to different vertices are related by the edges connecting the  vertices and, consequently share face areas. However, the connected faces generically have different shapes, e.g. a triangular face glued to a quadrilateral one. This geometric description is often referred to as a \emph{twisted geometry} \cite{FreidelSpeziale2010}. Remarkably, this brings together the spin network states of loop quantum gravity and the quantization of discrete geometry as described in Sec. \ref{sec:discrete_geometry} of this chapter. 

\subsection{Coherent states}  
\label{sec:coherent_states}
The states discussed in the previous section minimize and distribute the uncertainties in the non-commuting $P$-variables \eqref{qflux} coherently. Another interesting task, especially if the quantization of 4-dimensional spacetime geometries is the ultimate goal, is to balance uncertainties between the $P$-variables and the holonomies. While the spin network states minimize the fluctuations of the densitized triads, and leave the fluctuations of the holonomies maximal, it is desirable to find different states that balance these uncertainties. Such states are also loosely called semiclassical states. 

A proposal for these states that is well developed is that of  \emph{group coherent states} \cite{Thiemann:2000bv,Thiemann:2000bw,Thiemann:2000bx,Thiemann:2000by,Thiemann:2000ca,Bahr:2007xa,Zipfel:2015era}. It is based on coherent states on groups \cite{hall1994segal,hall2018coherent}. The basic idea is not to consider spin network states but infinite coherent linear combinations of such states, which increase the uncertainties of the $P$-variables, and reduce those of the holonomies. The main technical tool is to generalize standard coherent states  to ones on the quantization of phase spaces of the form $T^*G$ for $G$ a compact abelian group, in particular for $T^*$SU(2). A coherent state on SU(2) is an element of $L^2(\text{SU(2)})$ and takes the form
\begin{equation}
    \Psi_k^t(g) =\lim_{h\rightarrow k}e^{-t\, \Delta_\text{SU(2)}}\,\, \delta (gh^{-1}), \qquad k\in \text{SL(2,$\mathbb{C}$)}, 
\end{equation}
where $\Delta_\text{SU(2)}$ is the Laplacian on SU(2), $\delta$ is the group delta function peaked on the unit element, $k$ is an element of the complexification SU(2)$_\mathbb{C}=$SL(2,$\mathbb{C}$), and $t$ is a positive real parameter. An important property of the operator that acts on the Dirac delta on the right-hand side is that it maps it into an analytic function on SU(2). Thanks to analyticity, the function extends to SL(2,$\mathbb{C}$) and the variable $gh^{-1}$.  
One can show, \cite{Thiemann:2000bv,Thiemann:2000bw,Thiemann:2000bx}, that $\Psi_k^t(g)$ is peaked on a certain element $(h,t)$ of $T^*$SU(2), in the sense that $|\Psi|^2$ is peaked on $h$ and the expectation values of functions $f$ of invariant vector fields are close to $f(t)$, with small fluctuations.  Moreover, the parameter $t$ balances fluctuations between those of multiplication operators and those of the invariant derivatives in $L^2(\text{SU(2)})$. 

From these coherent states in  $L^2(\text{SU(2)})$, coherent states for loop quantum gravity can be constructed as product states
\begin{equation}
    \Psi_\gamma = \prod_{p\in\gamma} \Psi_{k_p}^{t_p}(h_p), 
\end{equation}
where the parameters $k_p, t_p$ are chosen judiciously, and adapted to the classical intrinsic and extrinsic geometry that is to be approximated \cite{Thiemann:2000bv,Thiemann:2000bw,Thiemann:2000bx,Thiemann:2000by,Thiemann:2000ca}. To enforce gauge invariance, these states have to be suitably adjusted \cite{Bahr:2007xa,Bahr:2007xn}.

These ideas can also be extended to graphs with infinitely many edges \cite{Sahlmann:2001nv,thiemann2001gauge}. This construction already leaves the Ashtekar-Lewandowski Hilbert space. Further generalizations of the same idea are  general complexifier coherent states \cite{Thiemann:2002vj} and r-Fock measures which we will discuss in section \ref{sec:other_phases}.

\subsection{Quantum-reduced gravity}  
\label{sec:quantum_reduced}
There is a class of spin-network states, we call them reduced, for which the action of the flux operators, as well as all the others expressed in terms of the fluxes, is greatly simplified. The associated approximation is that the spins carried by the edges are large 
$j_I \gg 1,$
and we consider the leading order in $j$'s only. 
The word `reduced' comes from a class of models of LQG in which one fixes  coordinates in $\Sigma$ and restricts the space of data $(A,E)$ so that the resulting metric tensor is diagonal and $E$ is suitably rotated at each point, see for example \cite{Alesci:2013xd,Alesci:2014rra}. 
The states of those models can be identified as states of full LQG, of the type mentioned above, and described more precisely below, with the caveat that they are not gauge invariant \cite{Makinen:2020rda}. 

A reduced spin-network is defined on a (topologically) cubic graph in $\Sigma$.  Taking advantage of that topology, the edges of the graph can be grouped into three classes corresponding to the would be (if $\Sigma$ were endowed with a flat geometry and the graph would be geometrically cubic) three directions, say $i=x,y,z$, and oriented accordingly. A one-to-one correspondence is introduced between the directions of the edges and orthonormal generators $\tau_i=\tau_x,\tau_y,\tau_z \in \mathfrak{su}(2)$. In the irreducible representation $V_{p_I}$ we choose an orthonormal basis of the eigenvectors of the spin operator ${D'}_{(j_I)}(\tau_{i_I})$, where $i_I=x,y$ or $z$ is the direction of $p_I$. We label the elements of the basis by the eigenvalues $m=-j_I,...,j_I$. Given that data, to every edge $p_I$ there is assigned an entry, either ${D_{(j_I)}}^{j_I}_{j_I}$ or ${D_{(j_I)}}^{-j_I}_{-j_I}$.  The corresponding cylindrical function is 
\begin{equation} 
\Psi(A) = \prod_{I}\sqrt{j_I(j_I+1)}{D_{(j_I)}}^{\pm J_I}_{\pm J_I}(h_{p_I}(A)),
\end{equation}
where either $++$ or $--$ is chosen independently on each edge $p_I$ of the graph and in an arbitrary way depending on the state. Notice, that the reduced spin-networks are not gauge invariant though. 

All the  operators of the geometry  introduced in this chapter  were written in terms of the vertex-edge spin operators  (\ref{J}).  Their action on the reduced states is very simple. Indeed, if $v$ is a vertex of a given reduced spin-network state, and $p_I$ is an edge that meets $v$ and has direction $x$, then 
\begin{equation}
\hat{J}_i^{v[p_I]}\Psi = \begin{cases}
			\frac{\pm j_{I}\hbar}{2}\Psi, & \text{if } i=x \text{ and } p_I \text{ is outgoing} \\
           \frac{\mp j_{I}\hbar}{2}\Psi, & \text{if } i=x \text{ and } p_I \text{ is incoming} \\
           O(\sqrt{j_I}) & \text{if } i=y,z .
		 \end{cases}
\end{equation}
Owing to these simplifications, the contribution from a vertex $v$ to the quantum volume element (\ref{volume})    is 
\begin{equation}
  \left(\frac{\kappa\gamma\hbar}{2}\right)^{\frac{3}{2}}\sqrt{(j_x^-  \pm j_x^+)(j_y^-  \pm j_y^+)(j_z^-  \pm j_z^+)}  + O(\sqrt{j}),
\end{equation}
where $j_i^\pm$ is the spin carried by an outgoing/incoming edge and the spins add if both magnetic numbers have the same sign. Incidentally, according to the reduced states, the unknown factor $a_0$ should be chosen to be $\frac{1}{2}$ as one of the distinguished possibilities mentioned above. 

The subspace of reduced states is preserved by the action of the holonomy operators of spins $s\ll  j$ (up to lower order terms in $j$). Specifically, 
\begin{align}
{D_{(j_{J})}}^{m}_{n}(h_{p_J}(A)) \Psi(A) = \prod_{I\not= J}\sqrt{j_I(j_I+1)}{D_{(j_I)}}^{\pm J_I}_{\pm J_I}(h_{p_I}(A))\nonumber\\
\begin{cases}
\sqrt{(j_J+m)(j_J+m+1)}{D_{(j_J+m)}}^{\pm J_I+m}_{\pm J_I+m}(h_{p_J}(A)) & \text{if } n=m\\
O(\frac{1}{\sqrt{j}}) & \text{if } n\not= m 
\end{cases}
\end{align}      
For the subspace of states spanned by cubic spin-networks, there is an alternative to the above-mentioned quantization of the curvature scalar.  In particular, the action of the curvature scalar operator has been computed on the reduced states in \cite{Lewandowski:2021iun,Lewandowski:2022xox}. 

\subsection{Other phases of quantum geometry}
\label{sec:other_phases}
\begin{figure}[t] 
    \centering
    \includegraphics[width=0.9\textwidth]{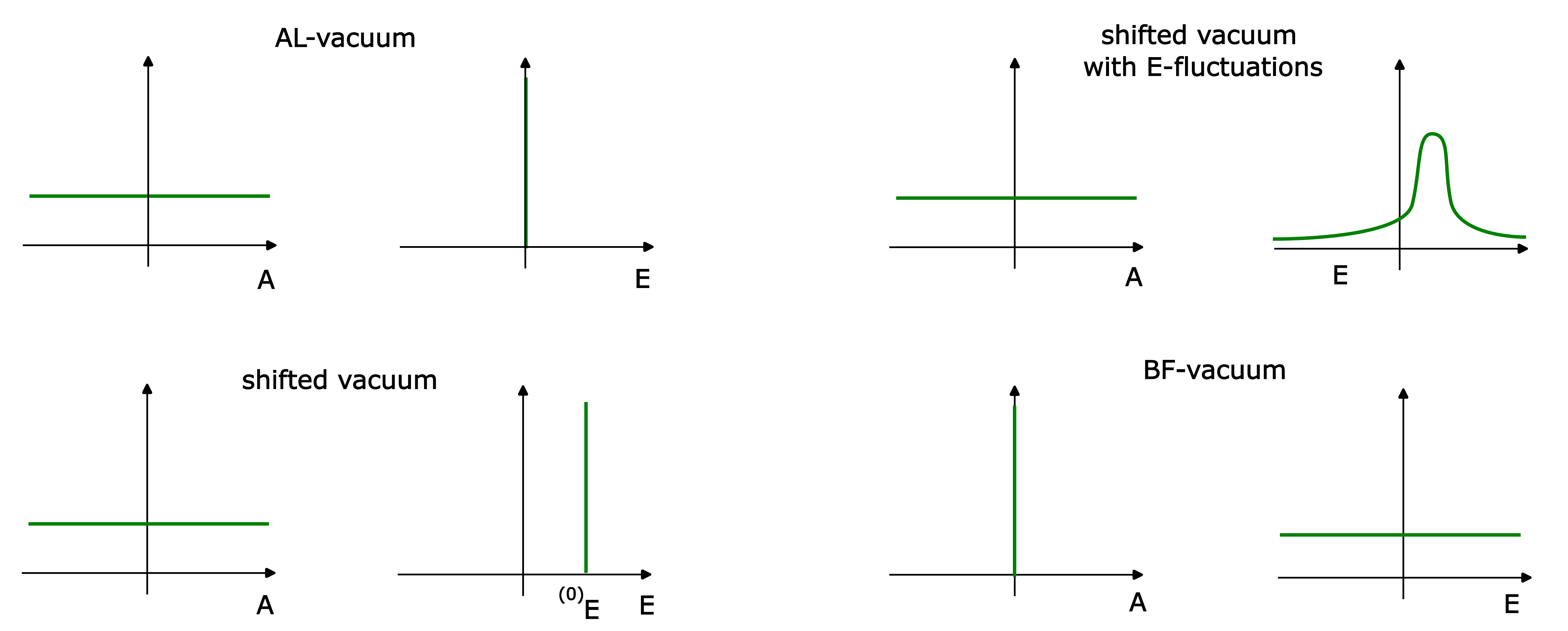} 
    \caption{Schematic depiction of the properties of various ground states of quantum geometry. Sketched is the behaviour in the $A$- and $E$-representation.}
    \label{fig:vacua}
 \end{figure}

In this section we will consider some extensions of the quantized geometry that we have described in the previous sections. We will proceed from straightforward extensions of the formalism to phases of quantum geometry that are less directly related. The properties of some associated ground states are sketched in Figure \ref{fig:vacua}.

The first extension is the introduction of matter. Scalar fields, gauge fields and fermionic fields can be coupled to gravity quantized as described in section \ref{sec:ALrep}. The resulting Hilbert space is a subset of the tensor product 
\begin{equation}\label{eq:structure_group_matter}
    \mathcal{H}=\mathcal{H}_\text{AL}\otimes \mathcal{H}_\text{matter}. 
\end{equation}
Details depend on the kind of matter that is coupled: Gauge fields with a compact structure group $G$ can also be quantized in terms of holonomies and fluxes, so effectively the only change is in the structure group, which becomes a tensor product $\text{SU(2)}\times G$, \cite{Corichi:1997us}. This does not change any of the features of the quantum geometry described in section \ref{sec:geometric_operators}. The same can be said of scalar fields \cite{Thiemann:1997rq}, which are quantized in analogy with gravity as described in section \ref{sec:ALrep}. 

A first change of the properties of the quantum geometry happens with the inclusion of fermions. This is because the fermion fields transform under the same gauge group as gravity. In loop quantum gravity, the fermions naturally appear as point like excitations \cite{Morales-Tecotl:1994rwu,Baez:1997bw,Thiemann:1997rq,Bojowald:2007nu}. To obtain gauge invariant states, excitations of gravity and fermions have to be coupled. For the quantum geometry, this means that the gravitational part can transform nontrivially under gauge transformations. Intertwiners at a vertex in the presence of a fermion must now also couple to the fermion. In the case of a Weyl fermion this means that an intertwiner $\iota$ lives in the space 
$
    \iota \in \text{Inv}(j_1\ldots, j_n, \tfrac{1}{2}). 
$
As explained in section \ref{sec:volume}, this changes the volume spectrum, sometimes dramatically. The volume operator is now non-trivial on 3-valent vertices that carry a fermionic excitation. For the intertwiner $\iota \in \text{Inv}(\tfrac{1}{2},\tfrac{1}{2},\tfrac{1}{2},\tfrac{1}{2})$ one obtains
$
    |q|\,\iota = {\sqrt{3}\iota }/{4},
$
via \eqref{eq:spectrum_threevertex}, for example. 

Matter also allows the construction of new geometric operators that relate matter and geometry, see e.g. \cite{Lewandowski:2015xqa} for a scalar field and \cite{Mansuroglu:2020acg} for fermions. 

In the supersymmetric extension of loop quantum gravity \cite{Gambini:1995db,Ling:1999gn,Bodendorfer:2011hs,Bodendorfer:2011pb,Bodendorfer:2011pc,Eder:2020erq,Eder:2021rgt}, further types of matter have to be considered and quantized \cite{Bodendorfer:2011hs,Bodendorfer:2011pb,Bodendorfer:2011pc}. 
The supergroup $\text{OSp}(\mathcal{N}|2)$ replaces the direct product of \eqref{eq:structure_group_matter} \cite{Eder:2020erq,Eder:2021rgt}. 
The gauge transformations thus mix matter and gravity degrees of freedom. In a formulation partially preserving manifest supersymmetry, 
fermionic excitations cannot therefore be point-like, but also extend along edges in $\Sigma$. Details of these constructions can be found in Chapter IX.3 of this volume. 

An extension of the formalism of section \ref{sec:ALrep} comes in the form of different representations of the holonomy-flux algebra $\mathfrak{A}_{\text{HF}}$ described in section \ref{sec:quantization}. The Ashtekar-Lewandowski representation is fundamental to loop quantum gravity because it is the only possible representation of $\mathfrak{A}_{\text{HF}}$ that carries a unitary representation of spatial diffeomorphisms and has a diffeomorphism invariant cyclic vector. One can compare the situation to that of relativistic quantum field theory. For free fields, with specified mass and spin or helicity, there is only one positive energy representation that carries a unitary representation of the Poincare group, and contains a cyclic Poincare-invariant vector: the vacuum. This representation is certainly fundamental, but there are other representations that are interesting and relevant, such as those for QFT at a finite temperature. The same can be said for loop quantum gravity. As already pointed out in section \ref{sec:quantization}, there are several definitions of $\mathfrak{A}_{\text{HF}}$ that differ regarding the imposition of certain higher order commutation relations. This also affects the representation theory. Details can be found in \cite{stottmeister2013structural,Koslowski:2011vn,Campiglia:2013nva,Campiglia:2014hoa}. The following applies to the definition given in \cite{Koslowski:2011vn}. 

Given a classical background field $^{(0)}\!E$, one can define, \cite{Koslowski:2007kh,Sahlmann:2010hn,Koslowski:2011vn},
\begin{equation}
    \pi(P_{S,f})= \widehat{P}_{S,f}+ P_{S,f}(^{(0)}\!E)\, \mathbbm{1},
\end{equation}
where $\widehat{P}$ is the operator \eqref{qflux} in the Ashtekar-Lewandowski representation. The Hilbert space and the action of the holonomies are unchanged. This representation describes quantum excitations over a background geometry given by $^{(0)}\!E$. This becomes even more clear when considering geometric operators that can be defined by a similiar regularization procedure  as in the Ashtekar-Lewandowski representation \cite{Sahlmann:2010hn}. The volume operator for a spatial region $R$, for example, becomes
$    \widehat{V}(R)= \widehat{V}_{\text{AL}}(R) + ^{(0)}\!\!V(R), 
$
 where $\widehat{V}_{\text{AL}}$ is the volume operator in the Ashtekar-Lewandowski representation and $^{(0)}V(R)$ is the volume of $R$ in the background geometry. 

It is also possible to include nonzero background fluctuations, by including a suitable operator valued shift,
$
    \pi(P_{S,f})= \widehat{P}_{S,f}+ ^{(0)}\!\!\widehat{P}_{S,f},
$
which can, for example, induce Gaussian fluctuations of $P$ in the vacuum state \cite{Sahlmann:2019elx}.
In these constructions, the background geometry is fixed, it can not be changed by the elements of $\mathfrak{A}_{\text{HF}}$. This can be modified 
\cite{Varadarajan:2013lga,Campiglia:2013nva,Campiglia:2014hoa} by suitably enlarging $\mathfrak{A}_{\text{HF}}$ by elements of the form 
\begin{equation}
  \widehat{W}\left(^{(0)}\!E\right)=\exp\int {}^{(0)}\! E\cdot\widehat{A}
\end{equation}
(see \cite{stottmeister2013structural} for subtleties in the case of a non-abelian structure group). 
Considering representations, one finds that these elements can shift the background. For a schematic depiction of  the properties of these representations, see Fig.~\ref{fig:vacua}.

If one is willing to go away from the algebra $\mathfrak{A}_{\text{HF}}$ and to introduce additional structures, other constructions become possible. One possibility is a quantum theory based on a vacuum that has, in some sense, properties opposite to those of the Ashtekar-Lewandowski vacuum. The \emph{BF-vacuum} \cite{Gambini:1997fn,Dittrich:2014wda,Dittrich:2014wpa,Bahr:2015bra,Drobinski:2017kfm} is an eigenstate of holonomies with the eigenvalues of a flat connection ${}^{(0)}\!A$, that is,  
$h\,|0\rangle_{\text{BF}}= {}^{(0)}\!h \,|0\rangle_{\text{BF}}. 
$
The observables related to $E$ have maximum uncertainties, and act by creating singular, two-dimensional excitations over the flat connection ${}^{(0)}\!A$. The construction of this new vacuum necessitates the introduction of a new algebraic structure comprising holonomies and fluxes and based on a class of two-complexes and their duals. However, it has been shown \cite{Drobinski:2017kfm} that for the case of structure group U(1), the analog to the BF vacuum can be obtained in a representation of a continuum theory, without any  discretization. In that theory, the discreteness emerges only on the quantum level as a property of the spectrum of the quantum holonomy operators.

In a similar spirit, one can keep the algebra generated by the holonomy functionals untouched, but change the rest of the relations in $\mathfrak{A}_\text{HF}$ by introducing a more regular smearing of the fields $E$ (three additional integrations against a smooth kernel). One obtains a new algebra \cite{Varadarajan:1999it,Varadarajan:2001nm}, which remarkably is, in the Abelian case where SU(2) is replaced by U(1), isomorphic to the algebra underlying the Fock representation of the quantum electromagnetic field. It can be used to define new representations of the holonomy part of the U(1)-analog  of $\mathfrak{A}_\text{HF}$ in which the holonomies have Gaussian fluctuations, the \emph{r-Fock representations}. Closely related are constructions of new states with Gaussian fluctuations, the \emph{complexifier coherent states} \cite{Thiemann:2000bv,Thiemann:2000bw,Thiemann:2000bx,Thiemann:2002vj}.  
It is possible to extend some of these constructions  to the case of linearized gravity \cite{Varadarajan:2002ht}, scalar fields \cite{Ashtekar:2002vh}, and even non-abelian gauge fields \cite{Assanioussi:2022rkf}. Also, \emph{shadow states} in the Ashterkar-Lewandowski Hilbert space can be defined using the r-Fock representations \cite{Ashtekar:2001xp}, which retain some of the properties of these representations.  

A very interesting generalization of the formalism of loop quantum gravity is to quantum group valued connections \cite{Lewandowski:2008ye}. In this case, the algebra of cylindrical functions over the group SU(2) is replaced by a (non-commutative) algebra of functions over a compact quantum group. It turns out that the co-multiplication of the quantum group, together with a certain quantum group automorphism, is precisely the structure needed to define an algebra of cylindrical functions as an inductive limit in this case \cite{Lewandowski:2008ye}.

Another generalization is to equip the spin network states of geometry with further structure that allows to encode the topology of the spatial slice $\Sigma$ \cite{Duston:2011gk,Duston:2015xba, Villani:2021aph}. Using this \emph{topspin}-formalism, it might be possible to describe topology change and quantum superposition of topologies. 

A link between Noncommutative Geometry and Loop Quantum Gravity was established by introducing a semi-finite spectral triple over the space of connections \cite{Aastrup:2008wa}. The triple involves an algebra of holonomy loops and a Dirac type operator which resembles a global functional derivation operator.

Another possibility to extend the formalism, while changing some technical aspects, is \emph{group field theory} (GFT).\footnote{There is by now an extensive literature on various aspects of this theory. For an introduction with a view toward loop quantum gravity see \cite{Oriti:2013aqa}.} 
One possible application is as an analog of  many particle theory in geometry, here  the analog of the one-particle states are loop quantum gravity states of a single vertex. However, the graphs used to describe these states are usually not thought of as embedded in a manifold. Mathematically speaking, group field theory can be obtained from the quantization of certain field theories on groups, with the propagator and interaction terms describing valence of the vertices and details of the gluing. 

The Hilbert space $\mathcal{H}_\text{GFT}$ is by definition a Fock space. Geometric operators such as the area and volume operators (see sections \ref{sec:geometric_operators} and \ref{sec:discrete_geometry}) define one-particle operations that can be lifted to $\mathcal{H}_\text{GFT}$ via second quantization. Gauge invariance forces couplings between the states at different vertices that can elegantly create extended quantum geometries. Multi-particle operators can generate various dynamics on these geometries. 

 There were  also attempts to generalize the quantum representation of the holonomy  and flux variables to the  differentiable or smooth category. The break through was the extension of the integral (\ref{integral}) to the cylindrical functions defined by all the piece-wise smooth curves \cite{Baez:1995zx}. For that purpose a smooth generalization of an embedded graph was defined, namely a web, whose edges can intersect and overlap infinitely many times, to form tassels. The resulting measure part and quantum representation of the cylindrical function observables give a complete and exact theory \cite{Lewandowski:1999qr,Fleischhack:2003vk}.  The webs can be used to define a generalization of the spin-network states, with many properties of the spin-networks. However, the flux operators, the quantum area, and quantum volume are not well defined on the cylindrical functions that are given by  webs of infinitely many self intersections or intersections with the $2$-surfaces used to define a given  operator. In other words, the operators are not densely defined. 

Another interesting idea, to generalize the class of differentiability of diffeomorphisms are analytic diffeomorphisms beyond a finite number of points \cite{Fairbairn:2004qe}. 

Diffeomorphisms are gauge symmetries of classical GR, which we quantize. That is why, according to the canonical approach, the manifold and its diffeomorphisms should pass naturally and in a scarified form into quantum theory. However, there is a radical, all-combinatorial approach to LQG, according to which the manifold does not exist as a background structure, and its structure is born in some other way, for example, through the polyhedral interpretation of vertices \cite{Rovelli:2010qx}.

\section{Discrete geometry}
\label{sec:discrete_geometry}

There is a second, striking path to the quantum discreteness of geometrical operators. In this approach one begins already with a discrete geometry; a geometry describable by a finite number of degrees of freedom, and making these degrees of freedom into a dynamical system works to understand the quantization of discrete geometry directly. There is a wealth of explicit results in this approach ranging from a general description of the phase space of shapes for convex polyhedra to explicit values for the quanta of volume of the simplest grain of space, a quantum tetrahedron. 

In this section we review this approach, frequently illustrating the methods and results using the quantum tetrahedron.  The tetrahedral geometry and phase space, while the simplest, are already quite rich. However, we caution the reader that many aspects of the tetrahedral case are quite special. For example, the volume of a tetrahedron generates an integrable dynamical system, which is not the case for convex polyhedra with more facets. The full complexity of multifaceted polyhedra and higher dimensional polytopes, such as the 4-simplex, and of complexes built up by gluing polyhedra together are of great interest. Indeed, a major open challenge in this setting is to build richly interacting networks of these discrete geometries that can be shown to be approximated by continuum regions of spacetime. 

\subsection{Dynamical Discrete Geometries: Evolving Polyhedra}

In a static, weak field the metric of spacetime is 
\begin{equation}
ds^2 = -(1+2\Phi)dt^2 +(1-2\Phi)(dx^2+dy^2+dz^2). 
\end{equation}
In this limit we can introduce a Newtonian gravitational field, given by $\vec{g} = -\vec{\nabla} \Phi$, and in regions free of mass this field satisfies $\vec{\nabla } \cdot \vec{g} = 0.$ Using the divergence theorem and focusing on a region where the gravitational field can be taken to be constant we have 
\begin{equation}
\oint_{\mathcal{S}} \vec{g} \cdot \vec{da} = \vec{g} \cdot \oint_{\mathcal{S}} \vec{da} = 0. 
\end{equation}
This holds for any direction of the gravitational field $\vec{g}$ and, so, for any small region enclosed by a surface $\mathcal{S}$, the oriented area of its boundary satisfies the closure condition
\begin{equation}
\label{SurfInt}
\oint_{\mathcal{S}}  \vec{da}  = 0. 
\end{equation}
This is a fact about any spatial Euclidean geometry and will be quite central to what follows. It is interesting to see this fact emerging here as a consequence of the constant gravitational field or small region limit of the theory. The gravitational field is what determines the local inertial frames and this identity holds in every sufficiently small spatial region of spacetime. 

\begin{figure}[t] 
    \centering
    \includegraphics[width=0.35\textwidth]{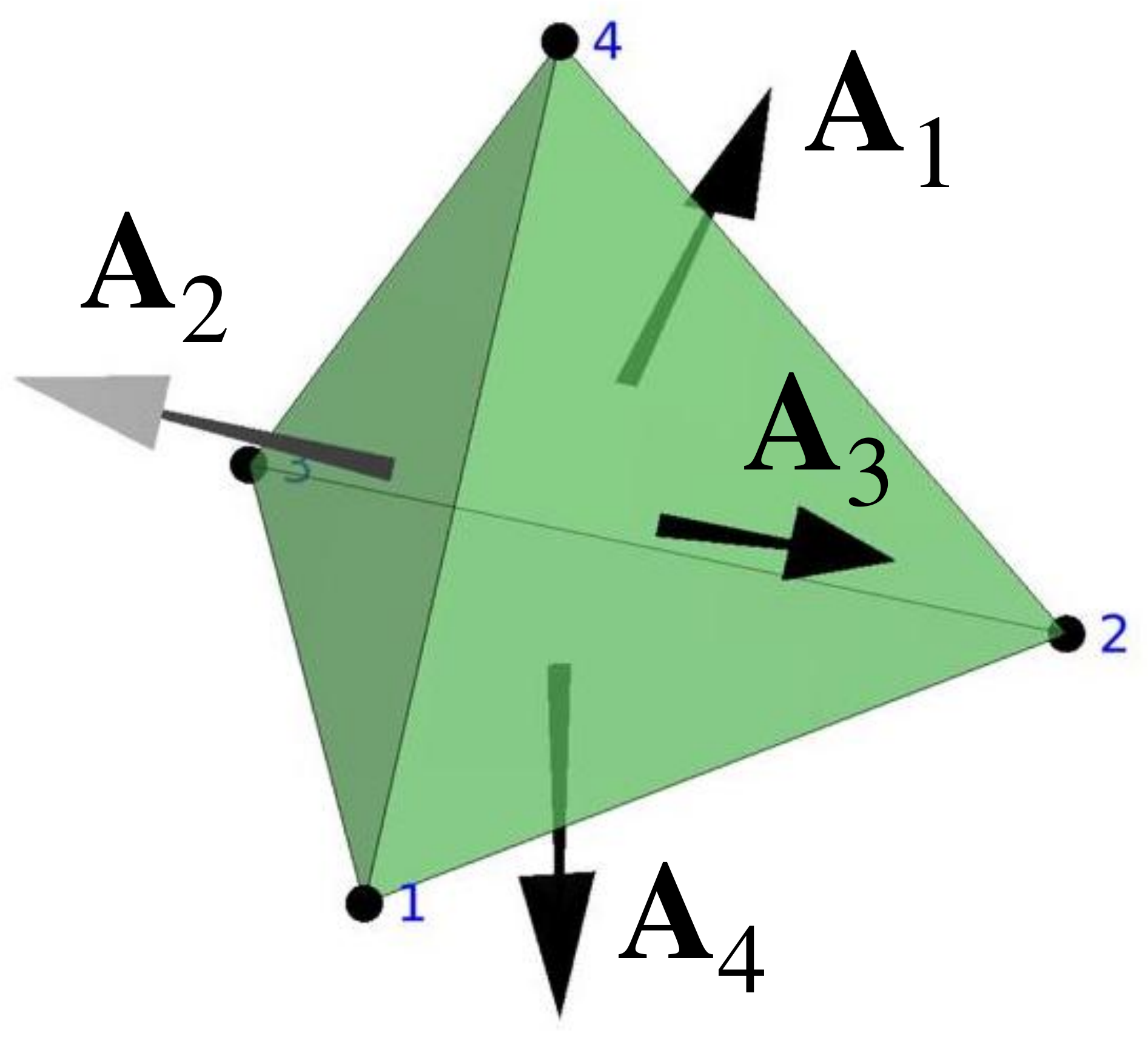} 
    \caption{A tetrahedron together with its area vectors. Figure created with GIBBON, \cite{Moerman2018}.}
    \label{ExamplePoly}
 \end{figure}

In the special case of a region that is a spatial polyhedron with $N$ facets, the integral \eqref{SurfInt} over the closed surface $\mathcal{S}$ breaks up into $N$ pieces each of which has a fixed direction $\hat{n}_\ell$ and one obtains
\begin{equation}
\label{DiscreteClosure}
\vec{A}_1 + \dots+ \vec{A}_N = 0,
\end{equation}
where each {\em area vector} $\vec{A}_\ell := A_\ell \hat{n}_\ell$ (no sum) points normal to the $\ell$th facet of the polyhedron and has magnitude $A_\ell$ equal to the area of this facet. (We use the index $\ell$ to label facets, and, of course, the area vectors $\vec{A}_\ell$ should not be confused with the Ashtekar connection $A_a^i$; the relation with the Ashtekar variables will emerge below.) Surprisingly, exactly the identity \eqref{DiscreteClosure} was leveraged by H. Minkowski, the same as of spacetime fame, to give a complete characterization of convex polyhedra at the close of the 19th century. Minkowski's theorem states that a set of $N$ vectors $\{\vec{A}_\ell\}_{\ell=1}^{N}$ satisfying the closure \eqref{DiscreteClosure} is in unique correspondence with a convex polyhedron of $N$ facets up to overall rotation in space \cite{Minkowski1897,BianchiDonaSpeziale}, see Figure~\ref{ExamplePoly} for an illustration of the case $N=4$.  

As we will see, the discrete closure condition \eqref{DiscreteClosure} is a powerful kinematical characterization of polyhedra. However, the true richness of the polyhedral approach emerges when we recognize the $\vec{A}_\ell$ as dynamical generators.  In accordance with Kepler's early insight, we will take each $\vec{A}_\ell$ to be an angular momentum, in the precise sense that they will be elements of the dual to the Lie algebra of $SU(2)$, $\vec{A}_\ell \in \mathfrak{su}^{*}(2) \cong \mathbb{R}^3$.  There are myriad roots for this choice, ranging from Penrose's introduction of spin networks \cite{Penrose1971,Penrose1972} to the fact that gravitational action integrals must always have the units of area. Here we will emphasize the fact that, with this choice, the closure \eqref{DiscreteClosure} becomes a gravitational Gauss law: it imposes a constraint on the kinematical variables $\{\vec{A}_\ell\}_{\ell=1}^{N}$ that determines the shape of the $N$-faceted polyhedron and in the same stroke it generates the overall rotational gauge symmetry that signifies that this shape is insensitive to its rotational orientation in space. Let us turn now to establishing this. 

We have emphasized that these angular momenta are in the dual to the Lie algebra because this is the deep reason that they come automatically equipped with a Poisson bracket, the Lie-Poisson bracket \cite{MarsdenRatiu}. The details of this construction need not distract us as this bracket, for each $\vec{A}_{\ell}$, is just the familiar one of angular momenta, often expressed in terms of the Levi-Civita $\epsilon_{ijk}$;  more generally, two functions of the full set of angular momenta, $f(\vec{A}_\ell)$ and $g(\vec{A}_\ell)$, have bracket given by the sum over facets of the familiar bracket:
\begin{equation}
\big\{f,g\big\}=\sum_{\ell=1}^{N}\, \vec{A}_\ell \cdot \left(\,\frac{\partial f}{\partial \vec{A}_\ell}\times \frac{\partial g}{\partial \vec{A}_\ell}\,\right),
\label{PB}
\end{equation}
where $\partial f/\partial \vec{A}$ is a convenient shorthand for $\partial f/\partial A^i \ (i=1,2,3)$. It is a quick check that for a fixed $\ell$ this gives $\{A_\ell^{i}, A_{\ell}^j\} = \epsilon^{ijk} A_\ell^k$. 

This bracket has the property that any finite sum of the $\vec{A}_\ell$ generates a rotation of those vectors in the sum around the axis of the resultant. For example, $A_{12} := |\vec{A}_1 +\vec{A}_2|$ generates rotations of $\vec{A}_1$ and $\vec{A}_2$ around $\vec{A}_{12}$ and leaves $\{\vec{A}_3, \dots, \vec{A}_N\}$ unchanged. More generally, $\hat{n}\cdot(\vec{A}_1 + \cdots + \vec{A}_N)$ is the diagonal generator of overall rotations of all of the vectors around the axis $\hat{n}$. This rotation does not, of course, change the closure of these vectors, Eq.~\eqref{DiscreteClosure}, or their relations to one another and, hence, leaves the shape of the polyhedron unchanged; this is the sense in which it generates a gauge transformation. The area vectors encode the intrinsic shape of the polyhedron and have a gauge symmetry with regard to its overall orientation in space. 

Summarizing, we have a remarkable triple of structures: (i) the geometry of any $N$-faceted polyhedron is described by the set of its area vectors $\{\vec{A}_\ell\}_{\ell =1}^N$ with each $\vec{A}_\ell \in \mathfrak{su}^*(2)$ and the full set satisfying the closure \eqref{DiscreteClosure}; (ii) these vectors have a Poisson bracket that can be used to generate dynamics; (iii) the magnitude of the closure relation \eqref{DiscreteClosure} doubles as a gauge generator that focuses interest on the rotationally invariant properties of the shape of the polyhedron. These structures allow us to study the Hamiltonian dynamics of polyhedra: simply choose any rotationally invariant function $H(\vec{A}_\ell)$ of the $\{\vec{A}_\ell\}_{\ell=1}^{N}$ and treat it as a Hamiltonian generator of the flow
\begin{equation}
df/d\tau = \{ f, H\}, 
\end{equation}
with $\tau$ a parameter along the flow. Rotationally invariant functions $H$ of the $\vec{A}_\ell$ (other than $|\vec{A}_1 + \cdots + \vec{A}_N|$) will typically generate flows that change the shape of the polyhedron, which is encoded in the angles between the various $\vec{A}_\ell$, but not the number of facets or the closure of the polyhedron. Hence we obtain a dynamics for the discrete geometries described by $N$-faceted polyhedra. 

We will call the Poisson space of $N$-faceted polyhedra that we have just described the \emph{space of polyhedra} $\mathcal{P}_N$,
\begin{equation}
\label{PoissonSpaceOfShapes}
 \mathcal{P}_N:=\big\{\vec{A}_\ell,\ell=1, \dots, N\,|\,\textstyle{\sum_\ell} \vec{A}_\ell=0\big\}/SO(3).
\end{equation}
This space is given by the product of $N$ copies of angular momentum space $\mathfrak{su}^{*}(2) \times \cdots \times \mathfrak{su}^{*}(2) =(\mathfrak{su}^{*}(2) )^N$, each copy isomorphic to $\mathbb{R}^3$, moded out by overall rotations of all of the vectors.

\subsection{The Phase Space of Polyhedra and Quantization}

Above we have exhibited the space of $N$-faceted polyhedra as a Poisson space. However, it is an immediate consequence of the definition of the Poisson bracket in Eq.~\eqref{PB} that each of the magnitudes $A_\ell := |\vec{A}_\ell|$ is a Casimir function of this bracket, that is, $\{A_\ell, f\} = 0$ for all $f$. Thus, this bracket always preserves the magnitudes of the $\vec{A}_\ell$. Indeed, a foundational theorem on the structure of Poisson manifolds says that they are foliated by symplectic leaves with each leaf labelled by the value of a Casimir \cite{Weinstein1983}. In our case, these leaves are the two-spheres picked out by the magnitudes $A_\ell$. Each of these leaves is a symplectic manifold endowed with the Kirillov-Kostant-Soriau symplectic form, which in this case is given by $\omega = A_\ell \sin \theta d\theta \wedge d\phi = A_\ell d\Omega$ with $(\theta, \phi)$ coordinates on the sphere and $d\Omega$ the solid angle, \cite{MarsdenRatiu,AquilantiEtAl2007}. Thus, each of the angular momentum spheres is a standard phase space and we can define a phase space of  shapes for polyhedra. 

Let $\Phi_N$ be the \emph{space of shapes of polyhedra} with $N$ facets of given areas $A_l$,
\begin{equation}
\label{SpaceOfShapes}
 {\Phi}_N=\big\{\vec{A}_\ell,\ell=1, \dots, N\,|\,\textstyle{\sum_\ell} \vec{A}_\ell=0,|\vec{A}_\ell |=A_\ell\big\}/SO(3).
\end{equation}
This space is given by the product of $N$ spheres $S^2~\times~\cdots~\times~S^2 = (S^2)^N$, obtained by fixing the magnitudes $A_1, \dots, A_N$, and moded out by overall rotations of all of the vectors. Its dimension is $\dim \Phi_N = 2N-6$, which is determined by the dimension of this collection of spheres, $2N$, minus three, for the conditions $\sum_\ell \vec{A}_\ell=0$, and minus three more for the division by overall rotations. This phase space was introduced and studied by M.~Kapovich and J.~J.~Millson in the somewhat different context of linkages \cite{KapovichMillson1996}. The advantage of this phase space is that the areas of the facets can be regarded as a fixed, parametric dependence during calculations. Nonetheless, we will freely transition back and forth between the Poisson and symplectic pictures, adopting whichever is more convenient for the discussion at hand. \\

\subsubsection{A first interlude on quantization: the quantization of area and spin network nodes and the meaning of quantizing grains of space}

Our first result on the quantization of discrete geometries follows immediately from our choice of dynamical variables, the $\vec{A} \in \mathfrak{su}^{*}(2)$. We associate a Hilbert space $\mathcal{H}_{j}$ with each facet of the polyhedron, or what is equivalent, with the edge of a spin network graph that crosses this face transversally. This is a carrier space for a unitary irrep of $SU(2)$, so that $\dim \mathcal{H}_{j} = 2j+1$. The $\vec{A}$ is a standard angular momentum variable and the quantization of its magnitude squared is given by 
\begin{equation}
\label{AreaSpectrum}
\hat{A}^2 |j m\rangle = (\hbar\kappa \gamma )^2 j(j+1) |j m\rangle, 
\end{equation}
where $|jm\rangle$ is a basis of $\mathcal{H}_j$; the physical scale of the quantization is the Planck area $a_{P} = \hbar \kappa$, and the gap between zero and the lowest lying area eigenstate has been parematerized by the Barbero-Immirzi parameter $\gamma>0$.  This remarkable result ties this choice of variables to a physical prediction: a Planck-scale discrete spectrum for the areas that connect neighboring regions of space. 

The $m$ quantum number in the states introduced here, the eigenvalue of $\hat{A}_z$, exhibits an orientation dependence of these states. This parallels the orientation dependence of any one of the area vectors $\vec{A}_\ell$. Only upon consideration of the full set of area vectors, satisfying the closure condition, is it that orientation independence is possible. Similarly, in the quantum case, the key to achieving orientation independence is to form a new state by combining facet states together using a rotationally invariant tensor that lives inside the product of the facet Hilbert spaces. Thus, we introduce the subspace of the tensor product Hilbert space $\mathcal{H}_{j_1}~\otimes~\cdots~\otimes~\mathcal{H}_{j_N}$ that is invariant under the global action of $SU(2)$ rotations  
\begin{equation}
\mathcal{H}_N = \mathrm{Inv}(\mathcal{H}_{j_1} \otimes \cdots \otimes \mathcal{H}_{j_N}).
\end{equation}
We call an invariant state, $|\iota \rangle$, of this space an {\em intertwiner} and $\mathcal{H}_N$ the space of intertwiners. Each state $|\iota \rangle \in \mathcal{H}_N$ can be expanded in a $|jm\rangle$ basis for each facet, 
\begin{equation}
|\iota \rangle = \sum_{m\text{'s}} \iota^{m_1 \cdots m_N} |j_1 m_1 \rangle \cdots |j_N m_N \rangle,
\end{equation}
and their components $\iota^{m_1 \cdots m_N}$ transform as a tensor under $SU(2)$ transformations,  cf. Sec.~\ref{sec:ALrep}. The defining condition that the states $|\iota \rangle$ are rotationally invariant can be expressed as the invariance of these components  under the diagonal action of $SU(2)$. 

Different choices of basis in the Hilbert space $\mathcal{H}_N$ emphasize different aspects of the resulting states; here we will focus on choices that highlight different aspects of the quantization of polyhedra \cite{BianchiDonaSpeziale}. That is, we will emphasize the geometry of quantum polyhedra. The literature explores a rich set of alternative choices, each with its own character and advantages \cite{GirelliLivine,FreidelSpeziale2010,FreidelZiprick2014,DupuisGirelliLivine}. The focus on quantum polyhedra here is for specificity and concreteness and is made to offer a route into this literature. That said, the reader will do well to remember that there are more ways of viewing the intertwiner space $\mathcal{H}_N$ than can be covered here.

The Hilbert space $\mathcal{H}_N$ is associated to an $N$-valent node of a spin network. Indeed, these are the gauge-invariant building blocks out of which the Hilbert space associated to a spin network graph $\Gamma$ is built. With this connection made, it is now possible to tie together the results of the first half of this Chapter to the discussion of this section. For simplicity, consider a single 4-valent node of a spin network, as discussed above this node corresponds classically to a  tetrahedron. The area vectors $\{\vec{A}_\ell\}_{\ell=1}^{4}$ discussed here are the fluxes $P_{S,f}$, defined at \eqref{Fl}, where the integration surfaces $S$ are each of the facets of tetrahedron. The function $f$, valued in the Lie algebra, gives the direction of the normal to the facet $\hat{n}_\ell$, and the rich non-commutativity of the components of each $\vec{A}_\ell$ can be seen as a consequence of the need to parallel transport the choice of internal frame at the center of the tetrahedron out to the facet along the path $p$ using the holonomies $h_p$, defined at \eqref{eq:holonomy}, or, alternatively, as a consequence of the gauge-invariant smearing along the facet using $f$, see \cite{freidel2013continuous} and \cite{cattaneo2017note}, respectively.  Thus, the $\vec{A}_\ell$ together with the closure \eqref{DiscreteClosure} and the Poisson structure \eqref{PB} are a discrete summary of the holonomy-flux algebra; parallel statements to all of these can be made about the quantum operators $\hat{\vec{A}}_\ell$ and $\widehat{P}_{S,f}, \  (\widehat{h}_p)^{a}{}_b $, defined at \eqref{HolFluxGen}. 
 
More generally, a picture of the quantum geometry of space is emerging: it can be seen as a collection of quantum polyhedra, invariant under local choices of frame, and glued along their equal area facets. Holonomies capture the curvature as collections of these polyhedra are traversed in closed paths. This rich, intuitive and mathematical picture is a consequence of the surprisingly simple quantization discussed above and is the foundation of the quantum geometry of loop quantum gravity. However, its interpretation is subtle and it is worth spending some time clarifying the conceptual content of this result.  

 The concreteness of the quantum polyhedral picture can be deceptive unless  emphasis is put on two aspects of the quantization procedure being considered: the truncation to a finite number of degrees of freedom of the gravitational field and the meaning of the modifier {\em quantum}. We take up each of these points in sequence. 
 
 \begin{figure}[t] 
    \centering
    \includegraphics[width=0.98\textwidth]{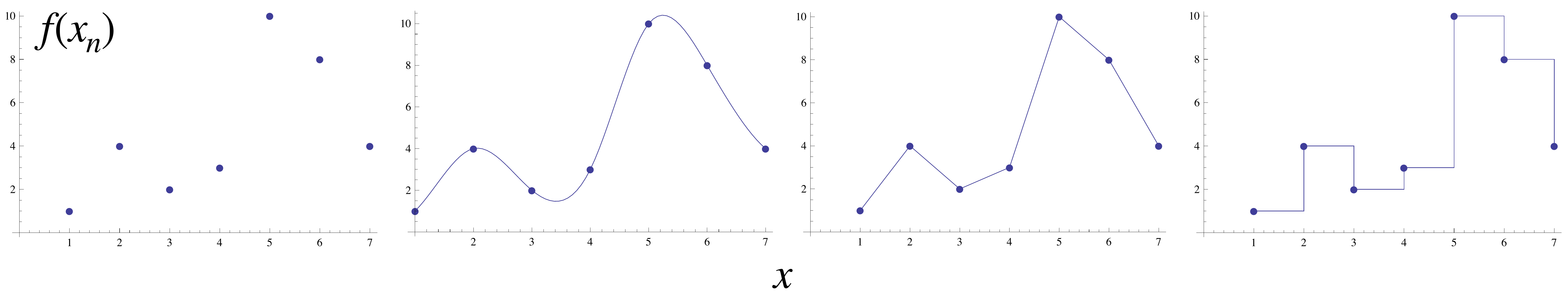} 
    \caption{A set of finite samples of a function $f(x)$ together with three different interpolations of the samples: polynomial, piecewise linear, and piecewise flat. These interpolation methods are analogous to mode sampling in cosmology, Regge geometries \cite{Regge1961}, and twisted geometries \cite{FreidelSpeziale2010}. }
    \label{Sampling}
 \end{figure}
 
 As emphasized in the preamble to this section, loop quantum gravity on a fixed graph $\Gamma$ is a truncation of the infinite number of degrees of freedom of the gravitational field down to a finite subset. Rovelli and Speziale, \cite{RovelliSpeziale2010}, have pointed out that this is analogous to sampling a function $f(x)$ at a finite number of points $f_n = f(x_n)$ for $n \in \{1, \dots, 7\}$, say. Given this finite sample it is not possible to reconstruct all of the information contained in the function $f(x)$, see Fig.~\ref{Sampling}, after \cite{RovelliSpeziale2010}. Nonetheless, much like in signal processing, there are definite insights that can be gained from various interpolations of these data. The polyhedra considered here are one such interpolation choice in quantum gravity; depending on how they are glued together they are like the piecewise linear interpolation of the 3rd panel, as in Regge calculus \cite{Regge1961}, or the piecewise flat interpolation of the 4th panel, as in twisted geometries \cite{FreidelSpeziale2010}.\footnote{It is the nature of these polyhedra as an interpolating scheme that leaves researchers unconcerned about the rigidity of the convexity assumption in the Minkowski theorem introduced above.} It is best not to confuse this interpolation choice with a fundamental statement about Nature. The claim in loop quantum gravity is not that there are little polyhedral pieces that make up space, but that the discrete geometry of polyhedra can model a finite number of the degrees of freedom of the gravitational field. And it is remarkable that, just as in the continuum, these degrees of freedom, the ones of discrete polyhedra, can be used to describe the dynamical system that is the evolving geometry of this approximation of a spatial region.  
 
 A result that must be emphasized in immediate counterpoint is that there {\em is} a claim of fundamental, physical discreteness here nonetheless. It is the spectral discreteness of the geometrical operators in loop quantum gravity. The operator that probes the area of a polyhedral facet, or, equivalently, of a spin network edge, has a discrete spectrum, Eq.~\eqref{AreaSpectrum}. Below we will see that the volume spectrum of a tetrahedron is also discrete. These spectra are a physical prediction of loop quantum gravity: were we able to experimentally probe Planck-scale regions of space the theory predicts that we would measure this discreteness directly. Thus, the meaning one should attach to the notion of a quantum grain of space is not that of a particular polyhedron (or any other model of part of the grain's degrees of freedom), but rather the insight that were one to measure aspects of the grain's geometry one would obtain spectral, quantum results. 

Another way of approaching this point, that also connects with the semiclassical methods discussed below, is to say that a quantum polyhedron is fuzzy. That is, we understand that when we refer to an energy eigenstate of an harmonic oscillator, this state is spread out over all the classical configurations that have this energy; the state does not correspond to a single, classical state, but is probabilistically spread out over all of them. In a semiclassical picture, the classical phase space of the harmonic oscillator is $\mathbb{R}^2$ with coordinates $(x,p)$ and symplectic $2$-form $\omega = dp \wedge dx$. The level set $H(x,p) = 1/2(p^2+x^2) = E$ is a circle and has the property that the pullback of the symplectic form to this set vanishes; we say that any submanifold of the phase space with this property is Lagrangian. The oscillator's energy eigenstates can be seen as quantizing these extended Lagrangian submanifolds of the phase space. Similarly, we will show below that a quantum polyhedron, e.g. a volume eigenstate of the tetrahedron, can be seen as spread out over the classical shapes that have that volume. In this sense, their shapes are quantum mechanically fuzzy. Before giving a detailed treatment of this perspective, let us complete our discussion of the phase space of shapes of polyhedra. 

\subsubsection{Action-angle coordinates}

 \begin{figure}[t] 
    \centering
    \includegraphics[width=0.75\textwidth]{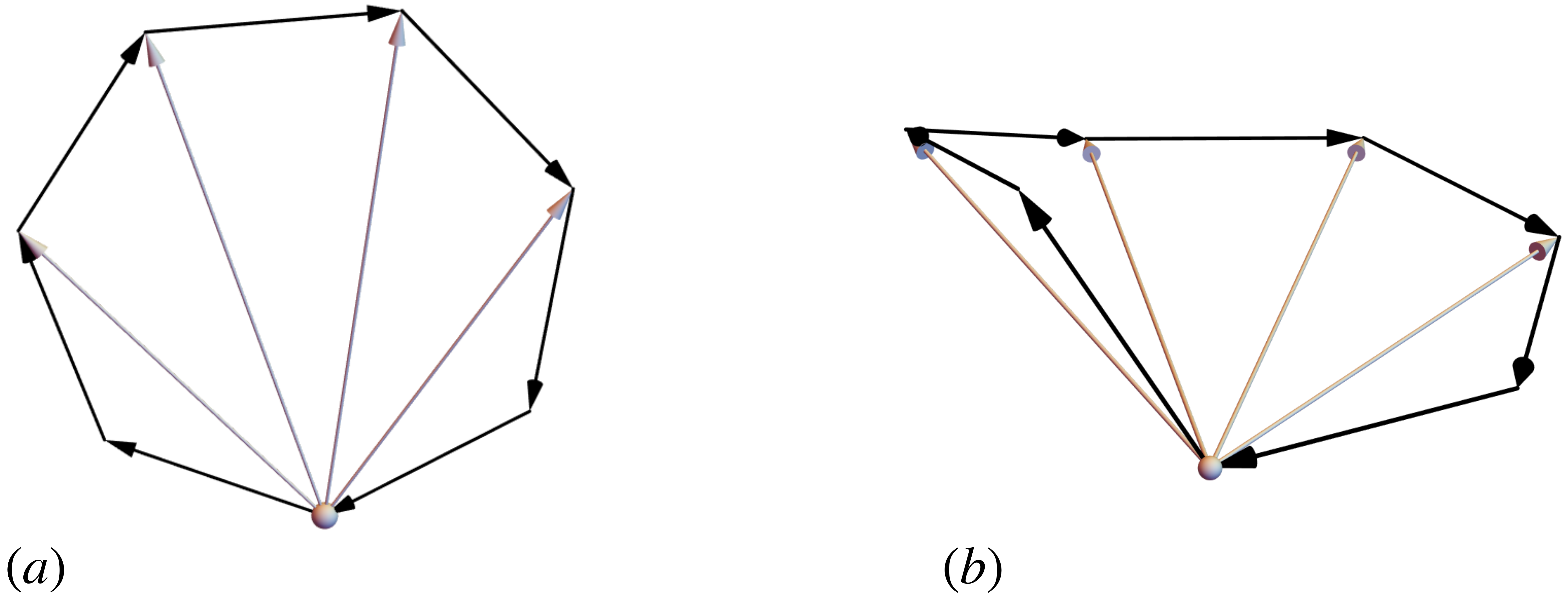} 
    \caption{(a) An example of a  planar configuration of seven area vectors, black arrows, satisfying closure. The magnitudes of the four gray vectors define the actions $\{I_k\}_{k=1}^4$ that generate the bending flows. (b) The result of following the $I_1$ flow $\pi/4$ radians and then the $I_2$ flow $\pi/6$ radians.}
    \label{Polygon}
 \end{figure}
 
One of the remarkable aspects of the space of shapes of polyhedra $\Phi_N$ that was recognized by Kapovich and Millson, \cite{KapovichMillson1996}, is that it comes naturally endowed with action-angle coordinates. (See, e.g. \cite{JoseSaletan}, for an introduction to this special class of coordinates on phase space.) On $\Phi_N$ these coordinates are most easily constructed geometrically. Fix the magnitudes of a set of area vectors $\{\vec{A}_\ell\}_{\ell=1}^{N}$ and consider a special configuration of these $N$ vectors in which they are all coplanar. This special configuration can be viewed as an $N$-sided polygon living in a plane of $\mathbb{R}^3$, see Fig.~\ref{Polygon}. Now, consider the set  of vectors $\vec{I}_k=\sum_{l=1}^{k+1} \vec{A}_l$, where $k=1,\dots, N-3$, that define a set of diagonals of this polygon.  We define the coordinate $\phi_k$ as the angle between the vectors $\vec{I}_k\times \vec{A}_{k+1}$ and $\vec{I}_k\times \vec{A}_{k+2}$, and the action variable $I_k=|\vec{I}_k |$ as the norm of the vector $\vec{I}_k$.

The action $I_k$ generates rotations of each area vector appearing in its definition around the corresponding diagonal $\vec{I}_k$. These  transformations are referred to as bending flows,  Fig.~\ref{Polygon}, panel $(b)$. In a slight generalization of the notion, we continue to refer to the resulting configuration of the area vectors as a polygon in $\mathbb{R}^3$, even though it is no longer planar.  Furthermore, up to overall rotations of all of the vectors, every polygon of $\mathbb{R}^3$, with the given values of the $\{A_\ell\}_{\ell=1}^N$, can be reached by varying the $I_k$ over their finite range of values and considering the bending flows they generate, \cite{KapovichMillson1996}. Thus, the set of $\{(\phi_k,I_k)\}_{k=1}^{N-3}$ pairs provide global coordinates on the $(2N$--$6)$-dimensional phase space $\Phi_N$. From~\eqref{PB}, it follows that these are canonically conjugate variables, $\{\phi_k,I_{k'}\}=\,\delta_{k k'}$, and so we have a set of action-angle variables. 

The simplest possible polyhedron, the  $N=4$ tetrahedron (Fig.~\ref{ExamplePoly}), to which we turn now, provides the perfect example for unpacking the details of all of the structures and correspondences that we have just reviewed. 

\begin{figure}[t] 
    \centering
    \includegraphics[width=0.98\textwidth]{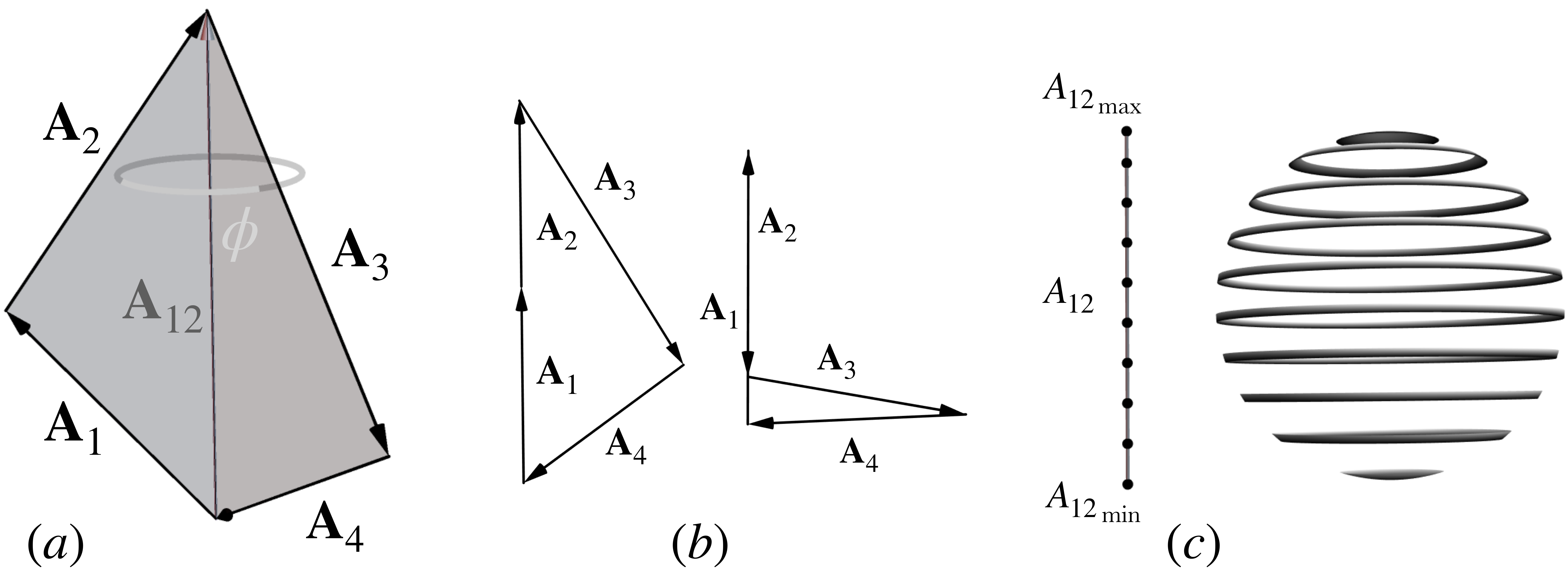} 
    \caption{(a) A non-planar configuration of the four area vectors of a tetrahedron. The action coordinate of the space of shapes $\Phi_4$ is $A_{12} =|\vec{A}_{12}|$. The conjugate angle, $\phi$, is given by the angle between the two planar wings in gray, spanned by $\{\vec{A}_{1},\vec{A}_{2}\}$ and by $\{\vec{A}_{3}, \vec{A}_4\}$, respectively. (b) The bending flow on the vector configurations that achieve the maximum and minimum values of $\vec{A}_{12} = \vec{A}_{1}+\vec{A}_2$ generates a rotationally equivalent configuration of vectors and hence corresponds to a single point of the phase space; the north and south pole of the phase space sphere depicted in the last panel. (c)  Each point of the interval of allowed values for the action $A_{12} \in [A_{12\text{min}}, A_{12\text{max}}]$ corresponds to a circle of rotationally distinct vector configurations, except at the endpoints of the interval. The collection of all of these circles and the single points at the extremal values gives the phase space of shapes of tetrahedra, which is topologically a sphere.}
    \label{TetQuad}
 \end{figure}
 
\subsection{Quantum Tetrahedra}
\label{sec:quantum_tet}

 \subsubsection{The Classical Geometry of the Space of Shapes of Tetrahedra}

The simplest discrete geometry that you can build in three-dimensional space is a tetrahedron. The action-angle variables and bending flow are particularly simple in this case: for $N=4$ the space of shapes, $\Phi_4$, is two-dimensional and its single action variable is $I_1:={A}_{12} = |\vec{A}_1+\vec{A}_2|$. This action generates the bending flow that increments the angle $\phi$ between the two wings of the quadrilateral formed by the closure of the four area vectors $\vec{A}_1, \dots, \vec{A}_4$, see the first panel of Fig.~\ref{TetQuad}. 

In fact, the action-angle coordinates help to uncover the topology of the space of shapes of tetrahedra $\Phi_4$. For a generic value of $A_{12}$, the bending flow sweeps out a full circle's worth of area vector configurations. Each of these configurations is distinct in the sense that no overall rotation of the four area vectors will transform one of these configurations into another. However, the action variable $A_{12}$ has a finite range,  spanning the interval $\max(|A_1-A_2|, |A_3-A_4|) \le A_{12} \le \min(A_1+A_2,A_3+A_4)$ and the end points of this range are special. In panel (b) of Fig.~\ref{TetQuad} the generic configurations of the area vectors at the ends of the range are illustrated. The bending flows of these special `flag' configurations no longer generate rotationally distinct configurations up to overall rotations of all four vectors. This means that at the extreme values of $A_{12}$ there is no longer a full circle of configurations, but only a single point. Thus, topologically, the space $\Phi_4$ is given by the Cartesian product of an interval with a circle that degenerates into a point at the two end points of the interval; this is precisely the topology of a sphere, Fig.~\ref{TetQuad}, panel (c).

\begin{figure}[t] 
    \centering
    \includegraphics[width=0.98\textwidth]{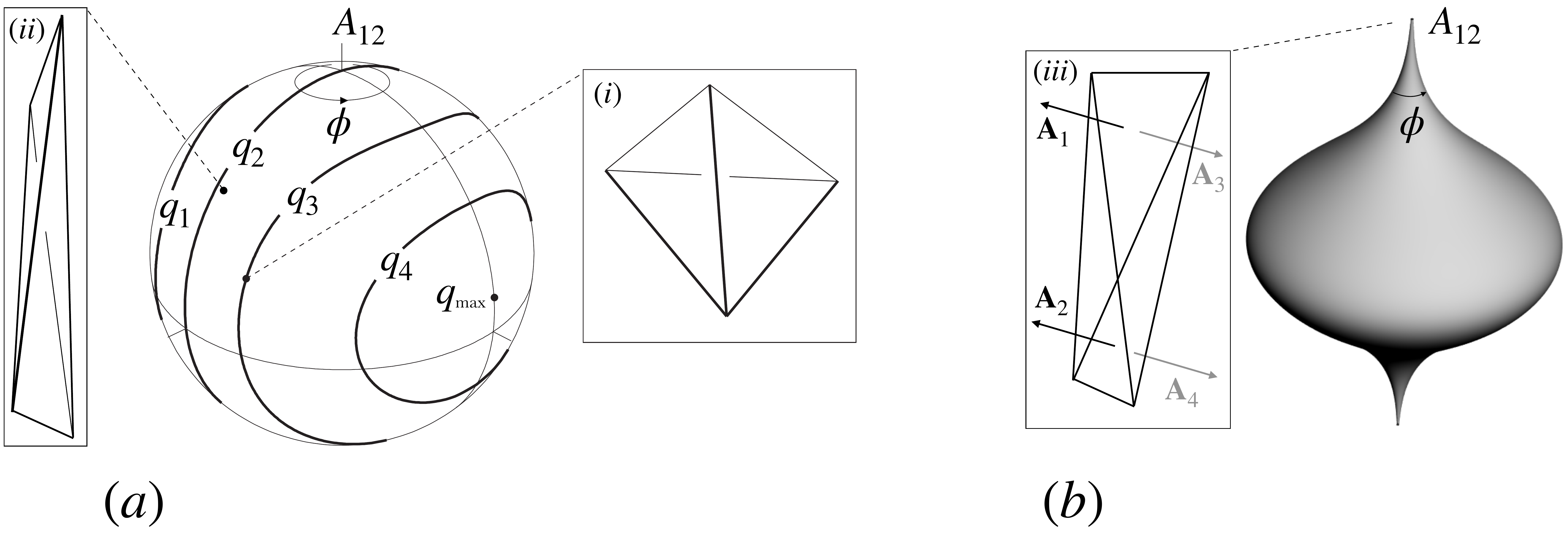} 
    \caption{$(a)$ The phase space of shapes of a tetrahedron $\Phi_4$. The inset $(i)$ shows a tetrahedron corresponding to a generic point of the space. The inset $(ii)$ shows a tetrahedron becoming pencil-like as the phase space point limits to the great circle $V=0$. The darkened contours illustrate quantized level sets, with values $q_i$, of the classical volume squared $Q = V^2$. $(b)$ When the area magnitudes are fixed to values that allow for planar tetrahedra, the space of shapes develops cusp singularities at the corresponding pole. Here the case where both $A_{12\text{min}}$ and $A_{12\text{max}}$ are singular is illustrated. The planar tetrahedral configuration at $A_{12\text{max}}$ is shown in inset $(iii)$.}
    \label{fig:SpaceOfShapes}
 \end{figure}

 Having discovered the topology and symplectic structure on the space of shapes of tetrahedra our main goal in this section will be to introduce a geometrically motivated choice of intertwiner on this space and to quantize it. This is a key step towards the development of the discrete-geometry path integrals for gravity called spin foam models, such as the Barret-Crane model \cite{barrett1998relativistic,barrett2000lorentzian}, and the EPRL model discussed in detail in Chapter IX.4. As stressed above, such an intertwiner should be a rotationally invariant operator that lives on the space of shapes. Geometrically a clearly motivated choice would be the volume of the tetrahedron \cite{Barbieri, BaezBarrett}. Remarkably, for a Euclidean tetrahedron the classical volume, $V$, also has a simple rotationally-invariant expression in terms of the area vectors
 \begin{equation}
 \label{Vol}
 V^2 = \frac{2}{9} \vec{A}_1 \cdot \left( \vec{A}_{2} \times \vec{A}_3\right). 
 \end{equation}
 This can be quickly checked by expressing the area vectors in terms of cross products of the vectors that run along the edges of the tetrahedron.   At first glance this expression seems to favor three of the four area vectors, but the closure relation, Eq.~\eqref{DiscreteClosure}, allows you to rewrite this formula in terms of any three. The left panel of Figure~\ref{fig:SpaceOfShapes} illustrates the space of shapes $\Phi_4$, a typical point corresponding to a tetrahedron in inset $(i)$, and some example contours of constant volume $V$.

Before proceeding to the quantization of this volume, there are three entwined subtleties about the classical phase space of shapes to clarify. First, consider area vectors that are coplanar, precisely the configurations used above to introduce the action-angle coordinates. These planar configurations of area vectors certainly lead to $V=0$, however, their interpretation is tricky. One might have expected that they all corresponded to planar configurations of the tetrahedron itself, like a tent squashed down onto its floor, but most of them do not. Observe that vectors pointing along the edges of a tetrahedron can be obtained by taking the cross-product of any two area vectors. For a planar configuration of the area vectors all of these cross-products are collinear and the tetrahedron has degenerated beyond planarity into a one-dimensional `pencil-like' tetrahedron. Note that, while these tetrahedra are degenerate, even a small deformation away from one of them is not and it is useful to keep these limit points in the phase space. Generically, the great circle on the spherical space of shapes obtained by merging the longitudes $\phi=0$ and $\phi=\pi$, which is all the configurations with $V=0$, consists completely of pencil-like tetrahedra. The exception is when the space of shapes includes planar tetrahedra, discussed in the third point below. 

Second, note that the bending flow is uninterrupted by planar configurations of the area vectors and continues past them. Area vector configurations symmetrically on either side of a planar configuration can be reached by sending $\vec{A}_\ell \mapsto -\vec{A}_\ell$ and performing an overall (gauge) rotation of all the vectors.  At the level of the tetrahedra, this maps outward pointing normals into inward normals. Since the area vectors are angular momenta, this is physically the time reversal of angular momenta, and the two hemispheres on either side of the  great circle $V=0$  can be seen as time-reversed tetrahedra in this sense. For many purposes then, including the quantization below, it is sufficient to work only with the hemisphere with outward pointing normals.  

Finally, the third subtlety is that the phase space of shapes becomes singular when planar configurations of tetrahedra are possible. Planar tetrahedra are only possible if the fixed area magnitudes satisfy an additional condition: either $A_1+A_2 = A_3+A_4$, or $A_1+A_2+A_3=A_4$, or a permutation of indices of one of these. The reason is that, for planar tetrahedra, the area vectors themselves cannot just be coplanar, but must be collinear, see inset $(iii)$ of Fig.~\ref{fig:SpaceOfShapes}. For collinear configurations of area vectors, the bending flow generated by $A_{12}$ is singular, and the phase space of shapes becomes cusped, panel $(b)$ of Fig.~\ref{fig:SpaceOfShapes}. The signature of these singular phase spaces is also imprinted on the quantization discussed next.

\subsubsection{Semiclassical Quantization of Tetrahedra}

Quantization of the tetrahedral volume formula \eqref{Vol} provides a physically compelling candidate for an intertwiner to use in determining gauge invariant states at the nodes of a 4-valent spin network. The clear difficulty with this expression is that, even in the well-adapted coordinates introduced here, it is cubic in the basic variables. This means that its quantization will inherently have operator ordering ambiguities and other subtleties. While the full spectrum is not known analytically, the spin network methods described in Section~\ref{sec:QuantumVol} have furnished explicit expressions for the matrix elements of the corresponding volume operator $\hat{q}$ in a particular basis, see, e.g.  \cite{BianchiHaggardVolLong}, for a summary of what is known. These methods have facilitated some analytic results and numerical investigations \cite{Thiemann:1996au,Brunnemann:2004xi,Giesel:2005bk,Giesel:2005bm,Brunnemann:2010yv,Schliemann:2013oka,Bianchi:2011ub,Brunnemann:2007ca}. A nice feature of the tetrahedral case is that the internal and external regularizations, discussed in Sec.~\ref{sec:QuantumVol}, agree in this case and the methods presented below give a compelling argument for fixing the arbitrary constant $a_0$ of Eq.~\eqref{volume} in this case. 

In this section we will review a second, semiclassical approach to the quantization of volume. This approach has yielded several explicit analytical results and is being actively pursued to extend existing results on spatial Euclidean tetrahedra to the case of Lorentzian tetrahedra embedded in spacetime.

The cubic character of the volume is not a problem for a semiclassical approach and leads into the interesting terrain of elliptic curves. The strategy is to fix the area magnitudes $\{A_\ell\}_{\ell=1}^4$ and to study the Hamiltonian evolution generated by $Q:= V^2 = \vec{A}_1 \cdot (\vec{A}_2 \times \vec{A}_3)$ on the phase space $\Phi_4$.  Taking $Q$ as a Hamiltonian, the brackets of Eq.~\eqref{PB} give the evolution of any function $f$ on $\Phi_4$ via the flow equation $df/d\lambda = \{f, Q\},$ with $\lambda$ the parameter of the volume flow, that is, $\{\lambda, Q\}=1$. 

The volume evolution on $\Phi_4$ can be expressed in a nice form in terms of the action angle coordinates $\{A_{12}, \phi\}$. Using the definition of $\phi$ as the angle between the two wings of the area vectors in panel (a) of Fig.~\ref{fig:SpaceOfShapes}, a cross-product calculation shows that the squared volume satisfies the relation
\begin{equation}
\label{ActionAngleQ}
Q A_{12} = 8/9 \Delta_{12} \Delta_{34} \sin \phi, 
\end{equation}
with $\Delta_{12}$ and $\Delta_{34}$ determined by Heron's formula
\begin{equation}
\begin{aligned}
\Delta_{12} &:= \frac{1}{4} \sqrt{[(A_1+A_2)^2-A_{12}^2][A_{12}^2 -(A_1^2-A_2^2)]},\\  \Delta_{34} &:= \frac{1}{4}\sqrt{[(A_3+A_4)^2-A_{12}^2][A_{12}^2 -(A_3^2-A_4^2)]}.
\end{aligned}
\end{equation}
The volume evolution of $A_{12}^2$ follows quickly, first note
\begin{equation}
d(A_{12}^2)/d\lambda = \{A_{12}^2, Q\} = 2A_{12} \{A_{12}, Q\} = 2 A_{12} \partial Q/\partial \phi, 
\end{equation}
where in the last equality the fact that $A_{12}$ and $\phi$ are conjugate has been used. Using the expression \eqref{ActionAngleQ} both to compute the partial derivative and to eliminate $\phi$ supplies the central result
\begin{equation}
\label{VolEllipticCurve}
d(A_{12}^2)/d\lambda = 1/9\, \sqrt{(4\Delta_{12})^2 (4 \Delta_{34})^2- 18 A_{12}^2 Q^2}.
\end{equation}
The argument of the square root is a quartic polynomial in the squared action $A_{12}^2$. 

This result demonstrates that the volume evolution is an elliptic curve. Let $x := A_{12}^2$ and $y:=dx/d\lambda$, then an elliptic curve in these variables is defined by an equality between $y^2$ and a cubic or quartic polynomial of $x$. Elliptic curves are famous in physics for the role that they play in the full, non-linear integration of the simple pendulum and in rigid body dynamics. The derivation above shows that the volume evolution is integrable. Indeed, explicit expressions for $A_{12}^2(\lambda)$ in terms of Jacobi's $\text{sn}(\lambda, m)$ can be found using this equation, see \cite{BianchiHaggardVolLong}. The function $A_{12}^2(\lambda)$ is periodic in $\lambda$ and is a parametrization of the darkened contours shown in panel (a) of Fig.~\ref{fig:SpaceOfShapes}. Of direct interest here is the fact that these curves can be used to provide a semiclassical quantization of the volume. 

The idea is to return to old quantum theory and implement the Bohr-Sommerfeld quantization rule. This rule states that the quantized values of an observable $Q$ can be found by requiring that the action integral, call it $I$, computed along a quantized level set of $Q$ should be equal to an half-integer multiple of Planck's constant
\begin{equation}
I(q) = \oint_{\gamma_q} A_{12} d\phi = 2\pi (n+1/2) \hbar,
\end{equation}
where $\gamma_q$ is the level-set contour $Q=q$ and $n$ is a nonnegative integer, $n \in \mathbb{Z}^*$. Once the action $I(q)$ is computed, the expression is inverted to obtain the quantized values $q_n$ of the quantized observable $\hat{q}$. 

\begin{figure}[t] 
   \centering
   \includegraphics[width=.92\textwidth]{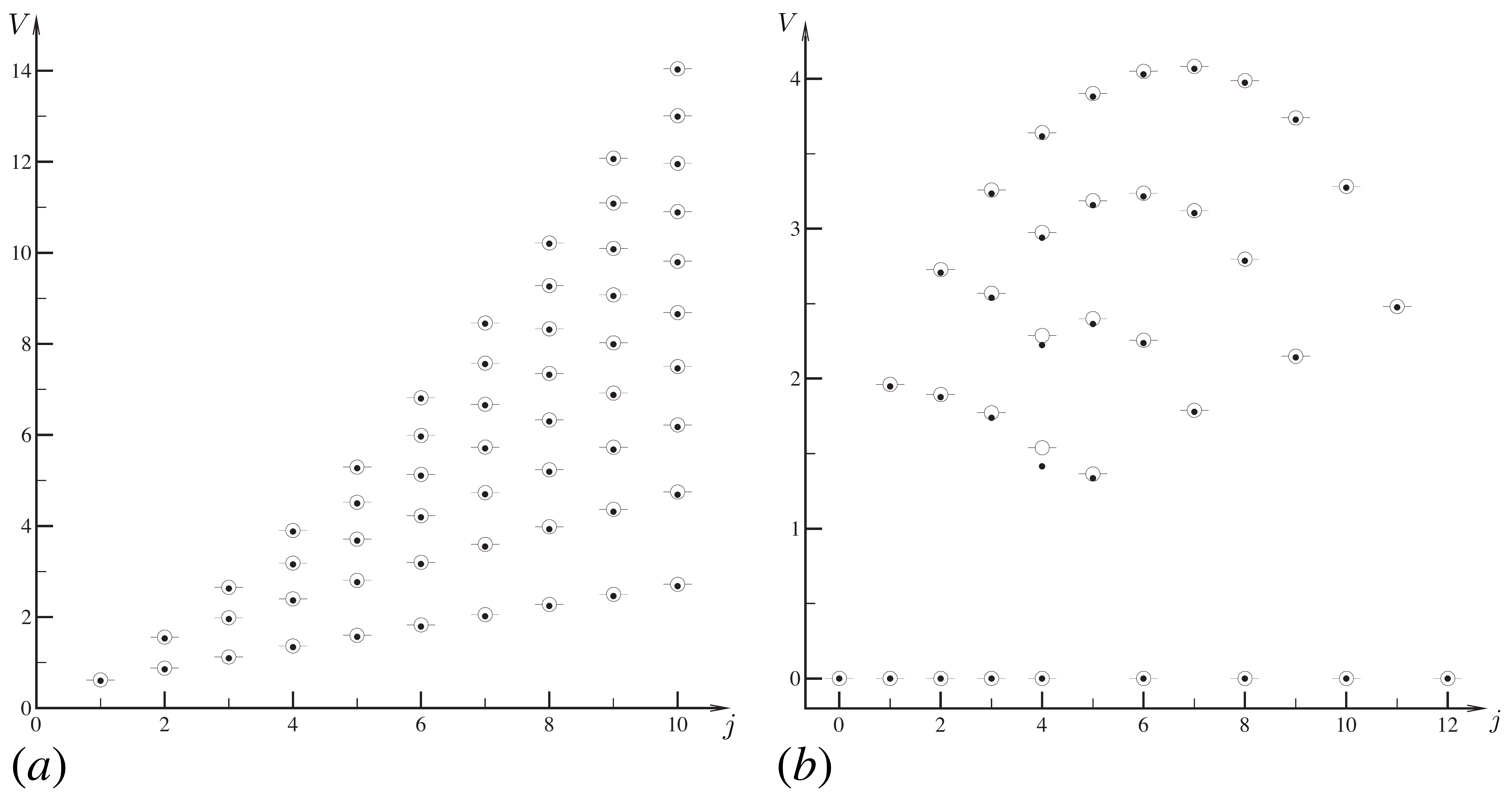} 
   \caption{ Comparison of the Bohr-Sommerfeld and Loop Quantum Gravity volume spectra. $(a)$ At left: a tetrahedron with area eigenvalues determined by $\{j, j, j, j + 1\}$. $(b)$ At right: a tetrahedron with spins $\{4, 4, 4, j\}$ and
$j$ varying in its allowed range. The Bohr-Sommerfeld values of the volume  are represented by black dots, the eigenvalues of the loop-gravity volume operator as open circles. Recall the spins $j$ and areas are related by $A_\ell =\sqrt{ j_\ell(j_\ell+1)} \hbar \kappa \gamma$; for these plots $\hbar \kappa \gamma =1$.}
   \label{SemiclassicalQuantization}
\end{figure}

This action integral can be carried out explicitly by parametrizing both $A_{12}$ and $\phi$ by $\lambda$,
\begin{equation}
I(q) = \Big(a K(m)-{\textstyle\sum_{i=1}^4} b_i \Pi(\alpha_i^2, m) \Big) q. 
\end{equation}
Here the result is expressed in terms of the complete elliptic integrals of the first kind, $K(m)$, and that of the third kind $\Pi(\alpha^2, m)$. The parameters $\{a, b_i\}$ and the moduli $\{m, \alpha^2_i\}$, both with $i\in \{1,2,3,4\}$, are functions of the $\{A_\ell\}_{\ell=1}^4$ and of $q$ through the roots of the quartic polynomial appearing under the square root in Eq.~\eqref{VolEllipticCurve}; their explicit expressions will not be needed here, but can be found in \cite{BianchiHaggardVolLong}. Using this expression, the Bohr-Sommerfeld values for the quantized level sets $q_n$ can be found. These Bohr-Sommerfeld values, black dots, are compared to the numerical diagonalization of the Loop Quantum Gravity spin network results, open circles, in Fig.~\ref{SemiclassicalQuantization}. The agreement, even at lowest order in $\hbar$, is striking.

\subsubsection{Perturbative/Non-perturbative Connections in Quantum Tetrahedra}

Using WKB theory the results above can be further extended to find analytic semiclassical expressions for the volume eigenstates  \cite{HaggardDissertation,HaggardILQGS2021}. Going beyond this, perhaps the most interesting application of these methods is currently under development \cite{AntuDoranHaggardForthcoming}. This has to do with the connection between these results and the theory of quantum curves, \cite{Gukov:2003na,Gukov:2011qp,Dimofte:2011gm,Norbury:2015lcn,Bouchard:2016uud}. Above we have seen that the volume flow traces out an algebraic curve, in this case an elliptic curve. All elliptic curves can be brought to Weierstrass normal form, where the relationship between the $x$ and $y$ variables introduced above can be written as $y^2 = 4x^3-g_2 x-g_3$ and the invariants $\{g_2,g_3\}$ classify the elliptic curve up to isomorphism. More generally, an algebraic curve is given by the vanishing of a polynomial in $x$ and $y$, $P(x,y)=0$. A quantum curve is a quantization of this relation where $x$ and $y$ are promoted to operators in the standard manner
\begin{equation}
\hat{x} = x, \qquad \hat{y} = \frac{\hbar}{i} \frac{d}{dx},
\end{equation}
and $P(x,y)$ is promoted to an operator $\hat{P}(\hat{x}, \hat{y};\hbar)$. The operator $\hat{P}$ is a differential operator in $x$, with coefficients that are polynomials in $x$, and, importantly, possibly power series in $\hbar$. States $\psi(x)$ annihilated by such an operator, $\hat{P}(\hat{x}, \hat{y};\hbar) \psi=0$, have turned out to be of central importance in knot theory, Chern-Simons theories, and matrix models. 

\begin{figure}[t] 
   \centering
   \includegraphics[width=0.98\textwidth]{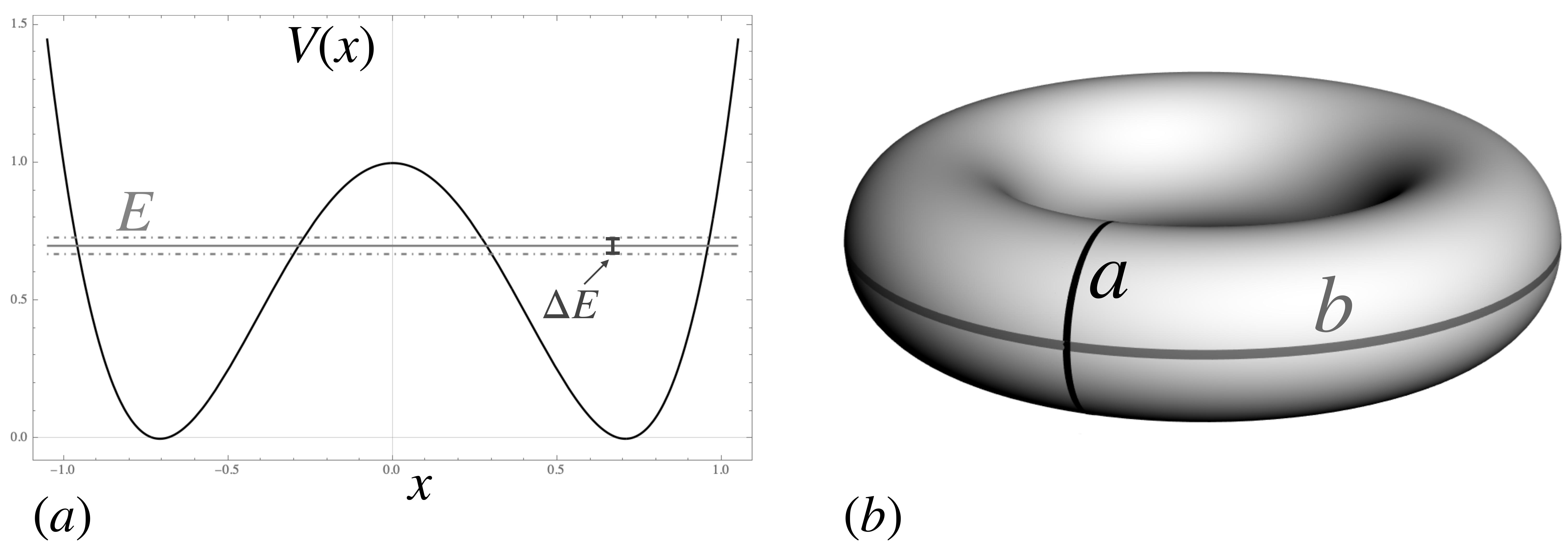} 
   \caption{(a) The potential energy $V(x)$ of a symmetric quartic oscillator with Hamiltonian $H(x,p)=p^2/2m+(2x^2-1)^2$. The level set $H=E$ is depicted by a solid gray curve. The quantum level splitting between the symmetric and antisymmetric energy eigenstates is illustrated as the gap $\Delta E$ between the dot-dashed horizontal lines. (b) The elliptic curves $p^2= 2m[E-(2 x^2-1)^2]$ for the quartic oscillator and \eqref{VolEllipticCurve} for the quantum tetrahedron can both be seen as genus-one Riemann surfaces. The action integrals along the topologically distinct $a$- and $b$-cycles are needed to compute the semiclassical approximations to the level quantization and non-perturbative tunneling of the quantum systems.}
   \label{fig:EllipticCurve}
\end{figure}

The case of elliptic curves is particularly interesting. In this case, Basar, Dunne, and \"Unsal have explained, in \cite{Basar:2017hpr}, an intriguing perturbative/non-perturbative connection using WKB theory. For concreteness consider an energy eigenstate of an anharmonic quartic oscillator. At the classical level a particle localized in the left well of the potential has an energy exactly degenerate with one localized in the right well, see panel (a) of Fig.~\ref{fig:EllipticCurve}. However, quantum mechanically this degeneracy is broken by non-perturbative tunneling between the two wells. The energy eigenstates have an exponentially small splitting $\Delta E$, shown schematically in the diagram by the dot-dashed curves. Surprisingly, the quantization of the energy level $E_n$ and the splitting between this level and its nearest neighbor, $\Delta E_n$, are not independent. The connection between these two quantities is explained by viewing the Hamiltonian level set $H=E$ as the elliptic curve $p^2= 2m[E-(2 x^2-1)^2]$. 

 As above, the Bohr-Sommerfeld quantization of the energy is computed by fixing the level set $H=E$ of the Hamiltonian and computing the action over a full period between, say, the two left-most turning points of the potential. Also interesting is the action integral connecting the two central turning points of the potential, which turns out to be what controls the level splitting $\Delta E_n$, \cite{Voros1983,Creagh1994}. The character of these two actions emerges more clearly if one complexifies the variables $x$ and $p$. Then, the elliptic curve $p^2= 2m[E-(2 x^2-1)^2]$ is the genus-one Riemann surface depicted in the second panel of Fig.~\ref{fig:EllipticCurve}. The two action integrals are simply the integrals of the action one-form, $pdx$, along the independent $a$- and $b$-cycles of this torus. Going  back to Riemann, the two periods, given by the energy derivative of these actions, were recognized to be related. In fact, these two periods are two independent solutions to what is known as a Picard-Fuchs equation. Basar, Dunne, and \"Unsal show that this relation extends to each order in $\hbar$, \cite{Basar:2017hpr}. This means that if you know the Bohr-Sommerfeld energy levels $E_n$ to a given order in $\hbar$ you can actually algorithmically construct their non-perturbative splitting $\Delta E_n$ to the same order in $\hbar$. 
 
 Many of the same mathematical structures are present in the volume quantization of the tetrahedron reviewed above. The elliptic curve corresponding to the volume, Eq.~\eqref{VolEllipticCurve}, can also be viewed as a complex, genus-one Riemann surface.  Recent work has uncovered the Picard-Fuchs equation for the periods of this family of Riemann surfaces, which are parameterized by the volume of the tetrahedron \cite{HaggardILQGS2021,AntuDoranHaggardForthcoming}. This work has established that one of these periods, the one along the $a$-cycle, is canonically associated with a real section of these Riemann surfaces; this period is the one associated to the action computed along the $Q=q$ contour that was used above to find the Bohr-Sommerfeld volume eigenvalues. Even more intriguingly, these researchers have shown that the second period is associated to a purely imaginary contour, the $b$-cycle, that can be put in correspondence with three-dimensional Lorentzian tetrahedra. These are tetrahedra embedded in a three-dimensional Minkowski space of signature $(-,+,+)$. These tetrahedra still satisfy a Minkowski theorem, but their squared volumes are negative. This work opens the way to a geometric and analytic understanding of non-perturbative tunneling between grains of space and grains of spacetime.

\subsection{Discrete Geometry Wrap-Up}

This section highlighted the emergence of Riemannian quantum geometry from its deep classical roots in Euclidean geometry. The same striking insight that Einstein brought to spacetime applies to discrete polyhedral geometries: they can be made into dynamical systems whose geometrical shapes evolve. Remarkably, these geometrical shapes can be quantized and have a clear correspondence with the spin network states of loop quantum gravity. This section showed how the states associated to the gauge-invariant nodes of a spin network can be viewed as quantum polyhedra and explored the case of the quantum tetrahedron in detail. Limitations of space precluded the treatment of other geometrical operators, but the discrete geometric approach also lends insight into the other operators introduced in Sec.~\ref{sec:geometric_operators}.

In particular, it was demonstrated how the area and volume of a quantum tetrahedron, which corresponds to a 4-valent node of a spin network, attain discrete spectra in a semiclassical approach to the quantization. Indeed these spectra give a complete set of quantum numbers for the 4-valent node. Higher valence nodes are tricky in more than one respect. For the 5-valent case, the polyhedral picture still supplies a clear proposal for a volume Hamiltonian \cite{Haggard2013}, but the dynamics generated by this Hamiltonian is chaotic \cite{Haggard2013,ColemanSmithMuller2013}. Even with more complicated volume spectra in hand there is the issue that with higher valence nodes, the dimension of the space of intertwiners grows and further quantum labels are needed. The action angle variables introduced above always provide a complete set of coordinates of the phase space and their is a corresponding set of quantum numbers, the angular momentum recoupling channels \cite{Yutsis:1962vcy}. While the recoupling channels are often used, another approach has also been developed that uses an enlarged Hilbert space where only the total bounding area of the quantum grain of space is fixed, this is termed the ${\rm U}(N)$ framework \cite{GirelliLivine,Freidel:2009ck,Freidel:2010tt,Borja:2010rc}. More general approaches looking for a full set of conserved charges at the boundary of spacetime are also under development \cite{Freidel:2015gpa,Freidel:2020xyx, Freidel:2020svx,Freidel:2020ayo}. 

There were many things that we could not cover in this section. One of the reasons to explain that the phase space develops singularities when planar tetrahedra arise, see Fig.~\ref{fig:SpaceOfShapes}, is that these are closely related to the role of asymptotic resurgence, \cite{fauvet2017resurgence,Dunne:2015eaa,Gu:2022fss}, in the study of quantum tetrahedra.  The planar configurations show up as logarithmic singularities of the elliptic functions of the third kind and contribute the zero-mode terms of the resurgent transseries \cite{AntuDoranHaggardForthcoming}. Another fascinating direction to go in is that of constant-curvature tetrahedra and 4-simplices. In three spacetime dimensions constant curvature tetrahedra quickly lead into the rich mathematical terrain of Poisson-Lie groups and quantum groups \cite{Bonzom:2014wva,Charles:2015lva,Dupuis:2020ndx,Bonzom:2022bpv}. The four-dimensional case of constant curvature simplices is deeply connected with the algebraic and quantum curves discussed above \cite{Haggard:2014xoa,Haggard:2015nat,Haggard:2015yda,Haggard:2015ima}. This work reawakened the long-standing exchange between Chern-Simons theory and loop gravity, \cite{Smolin:1995vq,Major:1995yz,Smolin:2002sz}, in particular studying connections to complex Chern-Simons theory \cite{Witten:2010cx,Dimofte:2014zga,Gukov:2016njj}. The work \cite{Haggard:2014xoa} pioneered the study of graph complement manifolds, generalizing the extensive work on knot complements \cite{kirk1993chern}.    

Throughout this chapter we have largely focused on spatial geometry. However, the discrete geometry approach does not stop there and can also be used to build up spacetime. This leads into the path integral approach of spin foam models \cite{Engle:2007uq,freidel2008new,Kaminski:2009fm,Perez:2012wv}, the subject of Chapter IX.4 of this Handbook. Indeed, viewing spin foam models as the gluing of discrete geometries has been productive and recently led to attempts to simplify the structure of these path integrals in models known as effective spin foams \cite{Asante:2020qpa,Asante:2020iwm,Asante:2021zzh}.

\section{Conclusion}

We offer a brief conclusion focused on drawing connections to other parts of this Handbook. The emergence of Riemannian quantum geometry informs and is informed by the developments discussed throughout this Handbook. The discreteness of quantum geometry has posed intriguing questions about spacetime entanglement, leading to a fruitful exchange with quantum information, the topic of Chapter IX.13, and to a long history of ideas in black hole physics, the subject of Chapters IX.9 and IX.10. While the focus of this chapter was on spatial geometries, the rich set of ideas around the Hamiltonian dynamics of these geometries is covered in Chapters IX.2 and IX.3, and the development of a path integral approach, in spin foam models, is covered in Chapter IX.4. The  computational methods necessary to calculate the amplitudes of a spin foam are covered in Chapter IX.5.  Enrichments of the notion of a spin network are covered in a Chapter IX.6 on Graphical Calculus and one on the Boundary Degrees of Freedom in Loop Quantum Gravity IX.12. Applications to Cosmology are discussed in Chapters IX.7 and IX.8, and the essential topic of the continuum limit of loop quantum gravity is taken up in Chapter IX.11. Finally, the volume concludes on the philosophical foundations of the theory in Chapter IX.14. 

\section{Acknowledgements}
H.M.H. was supported by grant no. 62312 from the John Templeton Foundation, as part of the \href{https://www.templeton.org/grant/the-quantum-information-structure-of-spacetime-qiss-second-phase}{`The Quantum Information Structure of Spacetime' Project (QISS)} and through the Perimeter Institute for Theoretical
Physics. Research at Perimeter Institute is supported in part by the Government of Canada through the Department of Innovation, Science and Economic Development Canada and by the Province of Ontario through the Ministry of Colleges and Universities.

JL was supported by grants of the the Polish Narodowe Centrum Nauki number  2018/29/B/ST2/01250 and number 2018/30/Q/ST2/00811. 

H.S. acknowledges the contribution of the COST Action CA18108.
\bibliographystyle{spphys}
\bibliography{references}

\end{document}